\documentclass[traditabstract]{aa}  

\usepackage{graphicx}
\usepackage{txfonts}

\usepackage{epsfig}
\usepackage{natbib}

   % ...

 \usepackage{psboxit}

                {\bfseries}%
                {}

\usepackage{psboxit}
 \PScommands
 \newenvironment{Note}%
                {\noindent\normalfont\normalsize
                 \begin{boxitpara}{box 0.85 setgray fill}%
                }%
                {\end{boxitpara}}

\newenvironment{erase}%
                {\noindent\normalfont\normalsize
                 \begin{boxitpara}{box 0.35 setgray fill}%
                }%
                {\end{boxitpara}}

\newcommand{\bnote}{\begin{Note}}
\newcommand{\enote}{\end{Note}}

\newcommand{\berase}{\begin{erase}}
\newcommand{\eerase}{\end{erase}}

 \usepackage{color}
 \usepackage{pstcol}

\begin{document}

\title{A revised radiometric calibration for the Hinode/EIS
instrument.} 

\author{G. Del Zanna}

\institute{DAMTP, Centre for Mathematical Sciences,  Wilberforce road Cambridge 
}

  \date{Received   ; accepted }

  \abstract{ A preliminary assessment of the in-flight radiometric calibration
of the Hinode EUV Imaging Spectrometer (EIS)  is presented.
This is done with the line ratio technique applied to a wide range of 
observations of the quiet Sun, active regions and flares from 2006 until 2012. 
The best diagnostic lines and the relevant atomic data are discussed in detail.
Radiances over the quiet Sun are also considered, with comparisons with 
previous measurements. 
Some departures in the shapes of the 
ground calibration responsivities are found at the start of the mission.
These shapes do not change significantly over time, with the  exception
of the shorter wavelengths  of the  EIS short-wavelength (SW) channel,
which shows some degradation. 
The sensitivity of the SW  channel at longer wavelengths
does not show significant degradation, while that of the 
 long-wavelength (LW) channel shows a significant degradation with time.
By the beginning of 2010 the responsivity of the 
LW channel was already a factor of two or more lower than 
the values measured on the ground. 
A first-order correction  is proposed. With this correction,
the main ratios of lines in the two channels become constant to within 
a relative 20\%, and the \ion{He}{ii} 256~\AA\ radiances over the quiet Sun
also  become constant over time.
This correction removes long-standing discrepancies for a number 
of lines and ions, in particular those involving the strongest 
\ion{Fe}{x}, \ion{Fe}{xiii}, \ion{Fe}{xiv},
  \ion{Fe}{xvii} and \ion{Fe}{xxiv} lines, where discrepancies 
of factors of more than two were found. 
These results  have important implications for various 
EIS science analyses, in particular for measurements of 
temperatures, emission measures and elemental abundances.
}

\maketitle

\section{Introduction }

The Hinode EUV Imaging Spectrometer 
(EIS, see \citealt{culhane_eis:07}) with its
two wavelength bands (SW: 166--212~\AA; LW: 245--291~\AA)
observes emission lines from a wide range of ions, allowing 
detailed measurements since December 2006 of electron densities and 
temperatures, as well as  emission measures and elemental abundances.
For such measurements an accurate radiometric calibration is 
of paramount importance. 
EIS was  radiometrically calibrated on the ground at RAL (UK) 
\citep{lang_etal:06}, providing an  overall uncertainty of about 20\%.
It has been generally assumed that relative line intensities are 
measured by EIS with greater accuracy.

A number of problems in the EIS intensities have been highlighted over the 
years, however, with typically factors of two discrepancies that could only
be ascribed to calibration problems. 
Significant (50\%) discrepancies were already found in 2007 August
observations of 
the strong \ion{Fe}{xiii} lines in the LW channel 
around 250~\AA\  \citep{delzanna:11_fe_13,delzanna:12_atlas}.
They could not be ascribed to problems such as either blending 
or the atomic data, given that excellent agreement 
with e.g. the \cite{malinovsky_heroux:73} [hereafter MH73] spectrum was found.
Significant (50\%) discrepancies were also found in 
a few of the strong \ion{Fe}{xiv}  lines \citep{delzanna:12_atlas}.
Again,  they could only be  ascribed to calibration  problems.

Significant  discrepancies in the 
\ion{Fe}{xvii} 204.6 and 254.9~\AA\  ratio  
 were reported by \cite{delzanna:08_bflare} 
and  \cite{delzanna_ishikawa:09}.  
These problems  are not present 
in previous (Skylab) observations nor in laboratory measurements,
where good agreement with theory is found (T. Watanabe, priv. comm.).
The  \ion{Fe}{xvii} lines form a branching ratio, for which atomic data 
are very accurate.

Significant  discrepancies in the two 
 strongest  flare diagnostic lines for EIS, the 
\ion{Fe}{xxiv} 192.0 and 255.1~\AA\ lines,
 were also reported by \cite{delzanna:08_bflare}.
These  discrepancies  are present in the literature, although often not specifically
mentioned. 
For example, \cite{hara_etal:11} presents a 2007 May 19 observation, where the
doublet ratio is approximately 2.9 photons, instead of the  expected 
(and well known) value of 1.85.
Recent flare observations in 2012 (discussed below) 
have presented discrepancies of more than a factor of two for both \ion{Fe}{xvii} 
and \ion{Fe}{xxiv} ratios,
suggesting that the  discrepancy in the relative calibration between the 
SW and LW channels  has increased over time.
Problems in this important ratio 
have recently  been confirmed by other authors 
(\citealt{young_etal:2013_flare}, H. Hara, priv. comm.).

\cite{wang_etal:11} performed a direct comparison between EIS SW 
and EUNIS quiet Sun [hereafter QS] observations in 2007, finding a small (20\%)
decrease compared to the ground calibration, a variation within 
the combined EUNIS and EIS uncertainties.
An estimate of the relative responsivity at a few wavelengths in the 
LW channel was also performed, finding again overall consistency,
but with larger uncertainties.

Starting in December 2006, regular observations of quiet Sun regions near
Sun center have been taken to monitor EIS sensitivity changes. 
The initial studies (SYNOP001 and SYNOP002) downloaded the full EIS spectral range.
However, after the X-band transmitter failure early in 2008, 
these were replaced with studies that only telemetred  a limited number of  
emission lines.
The radiances of the \ion{He}{ii} 256~\AA\ line show a clear decrease over time,
and a preliminary long-term correction for all the EIS wavelengths
was proposed,   assuming that 
the \ion{He}{ii} radiances should stay constant over time
(Mariska, priv. comm.).
This was implemented in the EIS software. 
The sensitivity decay was modeled as an exponential decay of the form 
$e^{-{t/ \tau}}$  where $t$ is the
time of the observation in days since the Hinode launch 
(2006 September 22 21:36 UT), and $\tau$
is the decay time, 1894 days (see EIS Software Notes \#1,2).

\cite{kamio_mariska:12} have recently presented 
QS radiances in  EIS lines obtained from these synoptic studies.
The EIS radiances were corrected for the sensitivity decay 
using an improved curve, still based on the \ion{He}{ii} radiances:
$(e^{-{t/ \tau_1}} + e^{-{t/ \tau_2}})/2$, where the e-folding times 
$\tau_1$ and $\tau_2$ are 467 and 11311 days respectively.
This correction was applied to both SW and LW channels, since
no evidence for a variations between the SW and LW channels was found.
The EIS radiances in lines formed above 1~MK show clear solar cycle
trends, and the \ion{He}{ii} radiances become constant, after the 
correction.
However, low-temperature lines such as \ion{Fe}{viii} 185.2~\AA\ and 
\ion{Si}{vii} 275.3~\AA\ show  unrealistic increases 
(see \citealt{delzanna_etal:10_cdscal, delzanna_andretta:11} for 
a discussion on solar cycle effects on spectral lines), a clear 
indication of a problem in the  correction.

The \ion{He}{ii} 256~\AA\ is the strongest LW  line in QS observations,
and should stay relatively constant, although 
as  shown in \cite{delzanna_etal:10_cdscal, delzanna_andretta:11}, the radiances of 
the helium lines are affected along the solar cycle  by the coronal radiation,
possibly because of photoionization-recombination effects
(see references in \citealt{andretta_etal:03}). Large variations between quiet Sun and
coronal hole areas are also present in helium lines.
So, the use of the \ion{He}{ii} 256~\AA\ for calibration is 
not ideal (see also below for further complications due to blending).
It should be accurate, however, if proper QS observations are selected
and the line carefully deblended.

\cite{mariska:12_eis_cal} re-analysed the set of synoptic observations,
noting that the radiances in the 
 \ion{Fe}{viii} 185.2~\AA, 
\ion{Si}{vii} 275.3~\AA, and \ion{Fe}{x} 184.5~\AA\ lines decreased in 
a similar way. The average decrease is small, of the order of 25\%, 
corresponding to an e-folding time of 7358 $\pm$1030 days.
The \ion{He}{ii} 256~\AA\ has a completely different behaviour, 
a nearly linear drop of a factor of 2 in the first two years of the 
mission, followed by a slower decay. The \ion{He}{ii} is severely
blended, but these blends cannot explain the difference. 
\cite{mariska:12_eis_cal} suggested that the \ion{He}{ii} should be
discarded, and that the results of the three above-mentioned lines indicate
a slow decrease of {\it both} SW and LW channels over time.

A more complete analysis of the in-flight changes of the 
EIS responsivities is in principle possible using line ratios
and not just line radiances, and the full EIS spectral range.
The aim of this paper is to present preliminary results from 
such an  analysis.  Similar  procedures adopted for the 
in-flight calibration of the SOHO CDS NIS and GIS channels 
\citep{delzanna01_cdscal,delzanna_etal:10_cdscal} are followed here.
\cite{delzanna01_cdscal} used the line ratio technique, described
in more detail below, while \cite{delzanna_etal:10_cdscal}
showed that the radiances of 
low transition-region lines are relatively unchanged 
along the solar cycle  and can be used to correct for the long-term
degradation (these corrections have been adopted by the CDS team 
to produce the final calibrated NIS data for the whole SOHO mission).

The present analysis has only been possible now that we have a more complete
understanding of the line identifications  and the atomic data.
Over the past ten years we \footnote{UK APAP network www.apap-network.org}
  have calculated  the atomic data for the main EIS lines  
 and have provided them to the community via the CHIANTI 
database\footnote{www.chiantidatabase.org} 
\citep{dere_etal:97,landi_etal:11_chianti_v7}.
Recently, we have performed large-scale R-matrix 
calculations for several ions to address the missing data problem
 of the  soft X-ray lines.  The calculations for 
\ion{Fe}{x} \citep{delzanna_etal:12_fe_10},
\ion{Fe}{xi} \citep{delzanna_storey:12_fe_11}, and 
\ion{Fe}{xii} \citep{delzanna_etal:12_fe_12}
are particularly relevant here because they resolved  
 long-standing discrepancies with observations for a few 
among the strongest EUV lines. These data 
 have been used in the present analysis and will be 
made available in a next CHIANTI release.

A review of line identifications and atomic data for all the EIS 
low-temperature lines was presented in \cite{delzanna:09_fe_7}.
A  review  of line identifications and atomic data for all the EIS coronal lines 
was presented in \cite{delzanna:12_atlas}.
The main flare lines were discussed in 
\cite{delzanna:08_bflare}, \cite{delzanna_etal:11_flare},  
and  \cite{delzanna_ishikawa:09}.  Several dozens of new lines 
have been identified, while similar numbers still await identification.
The majority of the EIS spectral lines, despite the high 
spectral resolution, turned out to be blended, and for this calibration
work a  
careful selection of the lines and observations needed to be done.

\section{The method and the data}
 
The method consists of choosing the appropriate observations and line ratios
to constrain the relative responsivities of the instrument
\citep{delzanna01_cdscal}.
The method has been applied in many instances, not only for the SOHO CDS. For example, 
 \cite{neupert_kastner:83} used  this method 
for an in-flight calibration 
of  the OSO V and OSO VII EUV spectrometers.
\cite{brosius_etal:98a} also used the same method to calibrate
observations from the 
 Solar EUV Rocket Telescope and Spectrometer 
 in 1995 (SERTS-95).
Line ratios were also used by 
\cite{young_etal:98}  to indicate problems in the calibration of 
 the SERTS-89 active region spectrum \citep{thomas_neupert:94}.

There are several good line ratios that can be used 
to check the in-flight relative calibration, as detailed below.
The best are branching ratios, with typical uncertainties of 
10\% or less. 
But there are also a number of ratios useful for the calibration,
those  that have small
variations with density and temperature.
Each ratio was assessed against various atomic calculations and
observations. The best calibrated EUV spectrum is that of MH73,
which shows typical agreement with theory within a remarkable 
few percent.

The SW and LW channels have a sufficient number of line ratios to 
check the shape of their effective areas.
The major difficulty and problem has been assessing the 
relative calibration between the two channels because
 very few line ratios are available. 
Another significant problem affect the use of data 
prior to 2008 Aug 24 (when the grating focus was adjusted).
Until this date, there
was an offset of about 2\arcsec\  in the pointing of the SW and LW
channels, meaning that the observations in the two channels were not simultaneous.
Since most observations have been made with the 1\arcsec\ slit,
it took more than twice the exposure time to observe the same region.
Most synoptic observations are single-slit observations, and there is 
no way to correct for this. Whenever raster observations have been available,
co-spatial regions have been selected for the present work. 
The lack of co-spatiality limits the use of data prior to August 2008, in particular for 
the flare \ion{Fe}{xvii} and \ion{Fe}{xxiv} lines, since 
 temporal variability is often significant on the exposure time scales.

A summary of the main line ratios chosen for the 
calibration is given in Table~\ref{tab:list_lines}.
A differential emission measure (DEM) analysis
(see \citealt{delzanna_thesis99} for the method used)
on many averaged spectra (on-disk, off-limb, over the
quiet Sun and active regions) has been performed, 
to assess possible known sources of blending.
Only  the relevant findings are summarised below.

\subsection{\ion{Fe}{viii} }

We have performed a preliminary assessment of the 
\ion{Fe}{viii} lines in terms of their use for 
calibration purposes. 
We used the atomic data (and identifications) for 
\ion{Fe}{viii} given in \cite{delzanna:09_fe_8}, and 
compared them with the recent large-scale atomic calculation for this ion
 performed by \cite{tayal:11_fe_8}.
Most predicted line intensities are very similar, within 10\%.
However, for a few lines  larger discrepancies
(30--50\%) were found. These discrepancies  originate from differences
in the atomic structure calculations of 
 \cite{tayal:11_fe_8} and \cite{delzanna:09_fe_8}, 
which reflect in different oscillator strengths, as shown in 
 \cite{tayal:11_fe_8}. The affected lines are originating from 
highly mixed levels, which include several strong lines 
observed by EIS, and that were the focus of the  \cite{delzanna:09_fe_8}
study. They are the 185.2, 186.6, 194.6, 197.3~\AA\ lines.
The large-scale  \cite{delzanna:09_fe_8} structure 
calculations provide energies in closer agreement to the 
experimental ones, and oscillator strengths in close 
agreement with the large-scale calculations of \cite{zeng_etal:03}.
It therefore appears that the \cite{tayal:11_fe_8} calculations
are not accurate for these lines.

However, for the 196.0~\AA\ line good agreement in the oscillator strengths
is found. 
The  185.2 / 196.0~\AA\ ratio is one of the most reliable.
We adopt a value of  5.3 photons
at log Ne [cm$^{-3}$] =9 and log Te [K]=5.65, obtained
with the  \cite{delzanna:09_fe_8} data, distributed within CHIANTI.


\subsection{ \ion{Fe}{ix}}

A few  \ion{Fe}{ix} from 
the 3s$^2$ 3p$^4$ 3d$^2$ configuration, identified by \cite{young:09},
could be used for the calibration, although with various limitations.
No branching ratios are available, and the lines
are close in wavelength.
Many of the lines were found blended during the present assessment. 
For example, the 188.4~\AA\ with \ion{Mn}{ix}, the 189.6~\AA  with \ion{Ar}{xi},
 and the 191.2~\AA\ line with \ion{S}{xi}.
All these blend are difficult to estimate.
We are left with the 189.9  and the 197.9~\AA\ lines.
The ratio of these lines is predicted to vary with both density and 
temperature by a significant amount, about 30\% in the temperature
interval where \ion{Fe}{ix} is expected to be formed, and 20\%
in the log Ne [cm$^{-3}$]=8--9. A value of 1 is assumed.

There is relatively good agreement between predicted
 \citep{storey_etal:02} and observed (using the the ground 
calibration) intensities for the EIS  \ion{Fe}{ix}  lines
\citep{young:09,delzanna:09_fe_7}, however the 
predicted intensities  for these lines are quite uncertain.
  The new scattering calculations for the other
iron ions mentioned in the introduction 
have shown that  cascading and resonance excitation due to the 
larger targets can affect both high- and low-energy  levels, 
by changing  the populations  up to 30--40\%. 
Work has started on  a similar large-scale calculation for 
\ion{Fe}{ix},  to improve  the current atomic data.

There is also the resonance 171~\AA\ line, which is however
barely visible in the EIS spectra, where the EIS sensitivity is about three 
orders of magnitude lower than at the peak around 195~\AA.
This line is also strongly density- and temperature-sensitive 
as shown in \cite{young:09} and \cite{delzanna:09_fe_7}.

\def\baselinestretch{1.}

\begin{table*}[!htbp]
\caption{Line ratios used for the EIS calibration.}
\begin{center} 
\footnotesize
\begin{tabular}{@{}lllllllllll@{}}

 \hline\hline \noalign{\smallskip}
Line ratio (\AA) & Predicted          & MH73 &  Observed  & $R_{\rm eff}$ & $R_{\rm eff}$ & Det. \\ 
                 & (log Ne=8--9)      &      &      DN/s  &  Pres.    & Ground     &     \\
\hline \noalign{\smallskip}

\ion{Fe}{viii} 185.2 / 196.0  & 4.5--5.3 (5.3 20\%)   &  & 1.23 (0.10,1.3/1.1) & 0.206 & 0.256 &  SW \\
\ion{Fe}{ix} 189.94 /  197.85 & 0.78--1.0 (1.0, 30\%)  &  & 0.68 (0.04,1.6/2.4) & 0.653 & 0.704 & SW \\

\ion{Fe}{x} 174.5 / 184.5 & 4.52--4.32 (4.4, 10\%) & 4.53(2\%) & 0.10 (0.01,0.3/3.4) & 0.021 & 0.024 & SW \\
\ion{Fe}{x} 177.2 / 184.5 & 2.6--2.49 (2.55, 10\%) & 2.53(0\%) & 0.16 (0.01,0.5/3.4) & 0.060 & 0.074 & SW \\

\ion{Fe}{x} 184.5 / 190.0 (bl)  & 2.94 (30\%) & (bl) &  0.92 (0.05,3.4/3.7)  & 0.304 & 0.332 & BR SW \\
\ion{Fe}{x} 207.45 / 184.5   &  0.14--0.18 (0.15, 30\%) &  & 0.05 (0.01,0.2/3.4) & 0.375 & 0.321 &  SW \\

\ion{Fe}{x} 257.3 (sbl) / 184.5 & 1.52--1.12    &  &  0.77 (0.08,7.5/9.8)  & 0.706 & 0.928 &  LW/SW \\

\ion{Fe}{xi} 188.2 / 192.8 (bl) & 4.8   &   &  2.1 (0.1,42.5/20)  & 0.427 & 0.559 & BR SW \\ 
\ion{Fe}{xi} 202.7 / 188.3 & 0.1 (30\%) &   & 0.06 (0.01,0.4/7.0) & 0.648 & 0.422 & BR SW \\ 
\ion{Fe}{xi} 178.1 / 182.2 & 0.27  &   &  0.07 (0.01,0.4/5.5)     & 0.254 & 0.231 & BR SW \\ 
\ion{Fe}{xi} 180.4 / 188.2 & 2.0 (20\%) &   & 0.34 (0.01,2.4/7.0) & 0.105 & 0.114 &  SW \\

\ion{Fe}{xi} 257.5 (sbl) / 188.2 & 0.165 &  &  0.05 (0.006,2.1/42.5)  & 0.411 & 0.413 & LW/SW \\

\ion{Fe}{xii} 192.4 / 195.1 (sbl) &  0.315 & 0.315(0\%) & 0.24 (0.01,5.7/23.4)  & 0.751 & 0.844 &  SW \\
\ion{Fe}{xii} 193.5 / 195.1 (sbl) &  0.67  & 0.67(0\%)  & 0.59 (0.01,13.9/23.4) & 0.873 & 0.924 &  SW \\
\ion{Fe}{xii} 186.9 (sbl) / 196.6 (bl) & 3.45--3.7 (3.5, 20\%) &  & 1.04 (0.11,1.5/1.5) & 0.282 & 0.381 & SW \\


\ion{Fe}{xiii} 209.9 (bl) / 202.0 & 0.15 (10\%) &   &  0.025(0.005,1.9/76.8)                & 0.169 & 0.159 &  BR SW \\
\ion{Fe}{xiii} 204.9 (bl) / 201.1 (bl) & 0.31 (30\%) &  0.27(-13\%)  & 0.088 (0.01,2.2/24.7) & 0.289 & 0.278 & BR SW \\
\ion{Fe}{xiii} 201.1 (bl) /197.4 & 5.0 (30\%) & 6.2(+24\%) & 2.65 (0.5,24.8/9.3)            & 0.540 & 0.416 &  BR SW \\
\ion{Fe}{xiii} 204.9 (bl) / 197.4 &  1.55 (30\%)  & 1.7(+10\%) & 0.23 (0.04,2.2/9.3)        & 0.154 & 0.115 &  BR SW \\
\ion{Fe}{xiii} 200.0 / 196.5 (bl) &  3.17--3.28 (3.3, 20\%) & 3.3(0\%) & 1.66 (0.3,13.1/7.8) & 0.512 & 0.576 &  SW  \\
\ion{Fe}{xiii} 209.6 / 200. & 0.74-0.685 (0.68, 20\%) &  & 0.05 (0.01,1.7/32.4)              & 0.072 & 0.081 & SW \\

\ion{Fe}{xiii} 246.2  / 251.9 & 0.51 (10\%) & 0.59(17\%) & 0.34 (0.01,2.9/8.4) & 0.651 & 0.635 & BR LW  \\

\ion{Fe}{xiii} 251.9 / 201.1 (bl) & 1.3(20\%)     &   &   &   &   &   LW/SW \\


\ion{Fe}{xiii} 251.9 / 204.9 & 4.2--4.8 &   &  2.73 (42.6/15.6) & 0.700 & 1.164 &   LW/SW \\

\ion{Fe}{xiv} 252.2 / 264.8 & 0.23 (10\%) & 0.25 & 0.08 (0.01,3.7/44.5)  & 0.331 & 0.421  &  BR LW \\

\ion{Fe}{xiv} 257.4 / 270.5 & 0.68 (10\%) &     & 0.32 (0.02,7.9/25.7) &    0.448 & 0.541       & BR  LW \\

\ion{Fe}{xiv} 289.1 / 274.2 &  0.065 (20\%) & 0.065 (0\%) & 0.013 (0.002,0.6/47.7)  & 0.211 & 0.232 & BR LW \\

\ion{Fe}{xiv} 274.2 / 211.3 & 0.69 (10\%)  & 0.77 (10\%) & 4.9(0.4,233/47.5)  & 9.225 & 9.636 &   LW/SW \\

\ion{Fe}{xiv} 270.5/ (264.7+274.2) & 0.26 (20\%) & 0.26 (0\%) & & & & LW \\

\ion{Fe}{xvi} 251 / 263 & 0.57 (20\%) & 0.66 (16\%) & 0.23 (0.02,3.9/16.6)  & 0.388 & 0.408 &  LW \\


\ion{Fe}{xvii}  254.9 / 204.7 & 1/0.93 (10\%)  &  & 0.64(0.06,5.6/8.7)  & 0.744 & 1.448 &  BR LW/SW \\

\ion{Fe}{xxiv} 255.1 / 192   & 1/1.85 (10\%)  &  &  0.067(0.007,3.5/52)  & 0.166 & 0.207 &  LW/SW \\


\ion{Si}{vii} 275.7 / 272.6 & 0.58 (10\%)   &  & 0.46 (0.07,0.1/0.3)  & 0.805 & 0.865 & BR LW \\

\ion{Si}{vii} 275.4 / 272.6 & 4.2--3.45  &  &  &  & & LW \\ 

\ion{Si}{x} 253.8 / 258.4 & 0.19 (10\%)    & 0.18(5\%)  & 0.13 (0.01,0.2/1.3) & 0.658 & 0.709 & BR  LW \\

\ion{Si}{x} 277.2 / 272.  & 0.84 (10\%)     &  0.87(2\%) & 0.53 (0.02,0.6/1.2) & 0.646 & 0.758 & BR LW \\

\ion{Si}{x} 261.1 / 277.2 & 1.37--1.35 (10\%) & 1.36(0\%) & 1.41 (0.08,0.9/0.6) & 0.969 & 0.926 &  LW \\

\ion{S}{x} 257.1 (bl) / 264.2 & 0.348 (10\%)  &   &       0.24 (0.03,0.2/0.9) &  & &  BR LW \\

 \ion{S}{xi} 285.6 (sbl) / 281.4 & 0.496 (10\%)  &    & 0.31 (0.05,0.3/1.1) & 0.634 & 0.635 & BR LW \\

\ion{Si}{vii} 275.4 / \ion{Fe}{viii} 185.2 &   & 0.63 &  0.61(0.06,2.6/4.3)  & 1.432 & 1.217 &  LW/SW \\

\noalign{\smallskip}\hline                                   
\end{tabular}
\normalsize
\tablefoot{The second column shows the theoretical ratios 
(photon units) within log Ne [cm$^{-3}$]=8--9 range, and 
optionally the chosen value, with an estimated uncertainty. The third column indicates the 
observed ratios from \cite{malinovsky_heroux:73} [MH73], and in brackets the 
percentage difference with the predicted ones.
The following column indicates the EIS  averaged ratio (in DN/s), with in 
brackets its variation (1 sigma) and the  averaged values of the two lines (again in DN/s).
The following two columns show the ratios of the effective areas $R_{\rm eff}$
obtained with the present calibration and the ground calibration. 
The last column shows the channel, and if the ratio is a branching ratio (BR).
 }
\end{center}
\label{tab:list_lines}
\end{table*}


\subsection{\ion{Fe}{x}}

The \ion{Fe}{x} identifications are summarised in 
\cite{delzanna_etal:04_fe_10}. The recent atomic calculations 
of \cite{delzanna_etal:12_fe_10} are adopted here.
The new data significantly affect the 257.26~\AA\ self-blend, the strongest
\ion{Fe}{x} line in the LW channel (identified in 
\citealt{delzanna_etal:04_fe_10}).
One problem is that the ratio of this self-blend with the 
lines in the SW band is both slightly density and temperature sensitive,
and is not ideal to check the SW/LW relative calibration.

The 174.5, 177.2, 184.5~\AA\ lines are all well observed in the SW channel.
Their ratios do not depend on  density or temperature, 
and excellent agreement (to within a few percent) with the MH73 observations is found.
QS observations  have been selected for these lines.
The 184.5 / 190~\AA\  is in principle  an excellent branching ratio, with 
both lines well observed. The  190~\AA\ line
is blended with at least a  known \ion{Fe}{xii} line (see below).
Further blending cannot be excluded, although the 
The \ion{Fe}{x} 184.5 vs. 190.0~\AA\ deblended branching ratio
shows a small scatter. The ratio has therefore been assigned a 
larger uncertainty (30\%).

The 207.45 / 184.5 ~\AA\ has a predicted ratio of about 0.15 photons using the 
\cite{delzanna_etal:12_fe_10} data. Note that there is a significant 
enhancement (factor of two) of the 207.45~\AA\ line, compared to the 
previous atomic calculations of \cite{delzanna_etal:04_fe_10}, due to the 
larger scattering calculation. On the other hand, the 
new calculations over-predict the 193.71~\AA\ line, which turns 
out to be very sensitive to the target employed, so it is not 
reliable for the calibration.

\subsection{\ion{Fe}{xi}}

\ion{Fe}{xi} produces a number of strong EIS lines.
New identifications of many of them was presented in 
\cite{delzanna:10_fe_11}, using the  atomic calculations of
\cite{delzanna_etal:10_fe_11}. 
 A few temperature-sensitive lines 
around  255~\AA\ were identified, in particular
 the self-blend at 257.5~\AA.
Excellent agreement with the MH73 data was found for a few \ion{Fe}{xi}
lines
\citep{delzanna:10_fe_11}, providing confidence in the calculations,
which resolved large (factors of 2-3) discrepancies.
 Here, the recent large-scale atomic calculations of 
\cite{delzanna_storey:12_fe_11} are adopted. They have improved the 
atomic data for the lines around  255~\AA.
For example,  Table~4 in this paper shows that the 
intensity of the 257.5~\AA\ line increases by 46\%. 
Such large increases were required to reduce the large 
disagreements between theory and EIS observations 
in the lines around 255~\AA, with these lines being 
far too weak compared to theory, as discussed in \citep{delzanna:10_fe_11}.
Further discrepancies were due to the use of the ground calibration,
as discussed below.

There are three good branching ratios in the SW channel. 
The 188.2 / 192.8~\AA\ one  is a complex,  as discussed 
at length in \cite{delzanna:10_fe_11}. 
The 188.2 is mainly due to \ion{Fe}{xi}, while the 
192.8~\AA\ is blended with a host of \ion{O}{v} lines,
 and in active regions with \ion{Ca}{xvii}, as described in 
\cite{young_etal:07a}.
The benchmark of a clean TR spectrum has shown good agreement 
between the 192.8~\AA\ and the 192.9~\AA\ \ion{O}{v} lines,
indicating that no further blending at 192.8~\AA\ with a cool line is present, as shown
in \cite{delzanna:09_fe_7}.
When off-limb observations (to avoid the presence of \ion{O}{v} lines)
are considered, there is a significant
and puzzling disagreement in this \ion{Fe}{xi} 
branching ratio (when the ground calibration is used).
One explanation would be the presence of a strong unidentified coronal line,
contributing about 30\% to the intensity of the observed line
 \citep{delzanna:10_fe_11}. 
There is a further \ion{Fe}{xi} line at 192.8~\AA, with a 
well-established branching ratio with the 188.3~\AA\ \citep{delzanna_etal:10_fe_11},
but this line is very weak. 
The deblended QS on-disk 188.2/192.8~\AA\ ratio is remarkably
constant in time, with a value of 2.0$\pm$0.1. If a blending 
with a coronal line was present, the unidentified line would have 
to have the same temperature of formation of \ion{Fe}{xi}, an 
unlikely coincidence. Considering the other calibration ratios,
we assume here that there is no unidentified line.
We adopt a QS off-limb value of 2.1 for this ratio. 

The \ion{Fe}{xi} 202.7 / 188.3~\AA\ branching ratio 
is more uncertain than other ratios, given the 
large range of values that are obtained with different 
atomic structure calculations.
The target adopted for the 
\cite{delzanna_etal:10_fe_11} scattering calculation
overestimates the 202.7~\AA\ line by about 30\%, according to the 
largest structure calculation (48CT), as shown in the same paper.
We adopt here  a branching ratio of 0.1, from the 48CT calculation.
The third  branching ratio,  178.1 / 182.2~\AA,  is only measurable in 
AR observation, due to the low signal in the first line.

The 180.4 / 188.2~\AA\ ratio is well established in terms of atomic
calculations \citep{delzanna_etal:10_fe_11}.
A conservative 20\% uncertainty is given to this ratio.
The 1--35 201.11~\AA\ transition \citep{delzanna:10_fe_11}
has increased excitation in the latest calculation, increasing 
its emissivity by 45\%.
A ratio of 0.032 (photons) with the strongest 180.4~\AA\ line
was adopted to deblend the \ion{Fe}{xiii} 201.1~\AA\ line.
The \ion{Fe}{xi} 257.5 / 188.2 is a good cross-channel ratio,
although it is slightly temperature-dependent. It varies by 
14\% in the log $T[K]$ =6.0--6.1 range.

\subsection{\ion{Fe}{xii}}

Several new identifications of important EIS 
\ion{Fe}{xii} lines were presented in \cite{delzanna_mason:05_fe_12},
in particular the self-blends at 186.9 and 195.1~\AA, 
used extensively to measure densities.
Here, we use the large-scale atomic calculations of  
\cite{delzanna_etal:12_fe_12}, where we obtained significant improvements 
 for the 186.9 and 196.5~\AA\ lines over our  previous calculations 
\citep{storey_etal:04}. The intensities of these lines are about 30\% 
higher, providing lower electron densities, now in agreement with those
obtained from other ions.

The 192.4, 193.5, 195.1~\AA\ are extremely strong in any 
observation and their ratios
are not sensitive to density and temperature (at high densities
the 195.1~\AA\ does change), so they are excellent for the calibration.
The MH73 observations are in exact agreement with theory.
A large piece of dust on the EIS CCD affects the 193.5~\AA\ line,
so the area was avoided.

The  186.9 / 196.6~\AA\ ratio is only slightly dependent on density and 
is used here. Note that the 196.6~\AA\ is slightly blended 
with a weak \ion{Fe}{viii} transition \citep{delzanna:09_fe_7}.
The QS observations were used for the calibration.

In principle, the 291/186.2~\AA\ ratio could be a good calibration ratio,
however the lines are weak and only measurable in AR spectra.
The  291~\AA\  measurement is  very uncertain because the line is at the very 
edge of the LW channel.  The weak 186.2~\AA\ line appears to be 
severely blended, not just with a known \ion{Fe}{xi} line.

There is a 190.07~\AA\ line blending \ion{Fe}{x} (see above).
Its intensity has been estimated from the observed 192.4~\AA\ line,
assuming a ratio of 0.023 photons. This ratio, with the 
\cite{delzanna_etal:12_fe_12}
atomic data,  ranges from 0.02 to 0.028  in the log Ne [cm$^{-3}$]=8--9 range.

\cite{delzanna_mason:05_fe_12} gave a tentative suggestion for the 
identification of the  strongest  decay from the 
 3s 3p$^3$ 3d configuration 
(from the $^4$D$_{7/2}$ 6--84 line), as a blend with 
another  \ion{Fe}{xii} line observed at 191.05~\AA.
However, Hinode/EIS spectra showed the 191.05~\AA\ line to be too
weak \citep{delzanna:12_atlas}.
The energy difference between observed and theoretical values for the 
lower 3s$^2$ 3p$^2$ 3d configuration suggests a wavelength around 188\AA.
There are a few possibilities, in particular that the line is blended with the 
strong \ion{Fe}{xi} 188.2, 188.3~\AA\ lines, or 
is the previously unidentified 
coronal line observed by \cite{delzanna:12_atlas}  at 188.37~\AA,
which has about the right intensity.
\cite{delzanna:12_atlas} suggested that this line could be blending the 
\ion{Fe}{x} 190.04~\AA\ line.
However, the \ion{Fe}{x} 184.5 vs. 190.0~\AA\ branching ratio
is very constant in time with a small scatter, which suggests 
that no significant blending of the 190.0~\AA\ line is present.
If the  \ion{Fe}{xii}  6--84 line intensity is deblended from 
the 190~\AA\ line, the \ion{Fe}{x} 184.5 vs. 190.0~\AA\ 
would increase over time instead of being constant (or decrease) at it should.

There is a weak 201.1~\AA\ line \citep{delzanna_mason:05_fe_12}
 blending the strong \ion{Fe}{xiii} transition. 
The intensity of the \ion{Fe}{xii} line has been estimated 
from the observed 191.05 line assuming a ratio of 0.24 (photons).


There are  other \ion{Fe}{xii} lines, but they are either weak,
strongly density-sensitive, or show significant disagreement with theory. 
For example, it has been suggested (P.Young, priv. comm.) that 
the 249.3~\AA\ could be useful. This line was identified in 
\cite{delzanna_mason:05_fe_12} as \ion{Fe}{xii}, 
however the observed intensity is significantly stronger than 
predicted. The $DEM$ modelling for the QS indicates a factor of 
two discrepancy, so this line is either blended, the identification
is incorrect, or the excitation rates for this line are incorrect.

\subsection{\ion{Fe}{xiii}}

There are several \ion{Fe}{xiii} lines useful for the calibration.
The new identifications presented in \cite{delzanna:11_fe_13} are adopted here,
together with the atomic  calculations of 
\cite{storey_zeippen:10, delzanna_storey:12_fe_13}.
The \ion{Fe}{xiii} lines are strong in off-limb  
QS and AR spectra, so only these observations have been considered.
The 246.2 and 251.95~\AA\ lines form a branching ratio, predicted at
0.51 (photons), in  excellent agreement with MH73 (0.59).

The observed intensities of the 246.2 and 251.95~\AA\
were  found  too low by a factor of at least 1.5
\citep{delzanna:12_atlas}, while they are in excellent agreement 
when the MH73 spectrum is considered.

The 209.9 / 202.0~\AA\ is also a good branching ratio,
although the 209.9~\AA\ line is slightly blended with an 
unidentified TR line  \citep{delzanna:09_fe_7}.
In on-disk AR observation, the average ratio in DN 
is 0.03. We adopt the off-limb 2007 Aug 19 value of 
0.025 for this ratio.

The 200.0 / 196.5~\AA\ is a ratio that is not much dependent on density,
so it is also used. 
Note that the  196.5~\AA\ line is blended
in  on-disk observations with an unidentified TR line \citep{delzanna:09_fe_7}.
We have use an off-limb observation for the 
200.0 / 196.5~\AA\ ratio.

The  251.95 / 201~\AA\ ratio  
varies only slightly with density and offers in principle 
a way to cross-calibrate the SW and LW channels. 
However, as pointed out in \cite{delzanna:11_fe_13},
the atomic calculations for the 197.4, 201.1, and 204.9~\AA\
lines, all originating from the same upper level, provide 
 branching ratios somewhat more uncertain than in other cases.

The 197.4~\AA\ line is relatively weak and has 
other lines nearby, so extreme care is needed to measure it.
The 201.1~\AA\ transition is considerably blended
(up to 15\% in AR spectra) with two known 
transitions from \ion{Fe}{xi} and \ion{Fe}{xii}, described in the 
relative sections.  The 201.1/197.4~\AA\ branching ratio 
is lower in off-limb AR spectra, an indication of further 
blending with a lower-temperature line. 
The averaged ratio in AR on-disk observations is 
3.1$\pm$0.5, while in the off-limb 2007 Aug 19 observation
was 2.65, a value adopted here.

The 201.1/ 197.4~\AA\ branching ratio has a theoretical 
value of 5.0 using the \cite{delzanna:11_fe_13} data.
 MH73 observed a ratio of 6.2, in 
quite good agreement once the blending in the 201.1\AA\ line
is taken into account.  
The 204.9 / 197.4~\AA\ theoretical branching ratio is 1.55, close
to the MH73 value of 1.7. Note that the 204.9~\AA\ 
is blended with a weak TR line, so the off-limb 2007 Aug 19 observation
is adopted here.

\subsection{\ion{Fe}{xiv}}

\ion{Fe}{xiv} lines are of particular importance for the calibration of the 
LW channel and the cross-calibration between the SW and LW.
Significant discrepancies in the LW line ratios and in the 
LW/SW one were found \citep{delzanna:12_atlas}.
Only AR observations are used here. In the QS, most of the 
\ion{Fe}{xiv} lines are hardly visible and become severely blended.
For example,  a new \ion{Fe}{xi} 
transition was  identified by \cite{delzanna:10_fe_11}
as blending the 264.7~\AA\ line.

We use the atomic data of \cite{storey_etal:00}.
\cite{liang_etal:10_fe_14} performed a larger scattering calculation
and found very similar (within a few percent) ratios, as described in 
that paper, with the exception of the  289/274~\AA\ branching ratio, for which 
\cite{liang_etal:10_fe_14} has 0.08, and 
\cite{storey_etal:00} 0.065 (photons). 
This ratio is sensitive to the target employed.
The latter value is adopted here, because it agrees with the 
MH73 value, obtained by deblending the 274~\AA\ line.
Note that the 274.2~\AA\ is  blended with 
a \ion{Si}{vii} transition that has a strong density sensitivity.
The intensity of this line has been roughly estimated here as
0.2 the intensity of the 275.3~\AA\ line.

The 252.2 / 264.8~\AA\ and 289.1 / 274.2~\AA\ are good
branching ratios, showing  excellent agreement with the MH73 spectrum.
The \ion{Fe}{xiv} 257.4 / 270.5~\AA\ branching ratio  is predicted to be
 0.68, but is often larger, a possible indication of a blend in the 
weaker 257.4~\AA\ line. 
The ratio of the  270.5/ (264.7+274.2) lines is predicted 
to be 0.26 (phot), exactly as observed by MH73.
The  274.2 / 211.3~\AA\  ratio  does not depend much on 
density and temperature. Good (within a relative 10\%)
agreement is found with the MH73 data.

\subsection{\ion{Fe}{xvi}, \ion{Fe}{xvii}, and \ion{Fe}{xxiv}}

\ion{Fe}{xvi} produces in active regions two strong lines,
which are not density-sensitive and offer in principle an excellent calibration ratio.
There are only three lines at the EIS wavelengths,
the strongest of which is the 262.98~\AA\ line.
We use the APAP atomic data of \cite{liang_etal:09_na-like}.
The \ion{Fe}{xvi}  265~\AA\ line  appears to be blended, while
the 251 / 263~\AA\ predicted ratio is not in very good agreement 
with the MH73 observations.

\ion{Fe}{xvii} lines in the EIS channels have been identified/benchmarked
in \cite{delzanna_ishikawa:09}. Here,  the atomic data of 
\cite{liang_badnell:10_ne-like} are used.
The two main lines for this ions, at 204.7 and 254.9~\AA,
 form  a  branching ratio.
As the \ion{Fe}{xxiv} 192 / 255.1\AA\ ratio, significant discrepancies
were found  by \cite{delzanna:08_bflare} 
and  \cite{delzanna_ishikawa:09}, with the LW lines being 
much weaker than expected.

The \ion{Fe}{xvii} 204.7~\AA\ line was found by 
\cite{delzanna_ishikawa:09} to be  blended 
  with a low-temperature line, identified
as partly due to \ion{Fe}{viii} \citep{delzanna:09_fe_8}.
However, the blending did not resolve the issue. 
Opacity effects would make the problem worse, so the only explanation
is a problem in the EIS calibration.

Flare observations of the right type are needed, to use 
\ion{Fe}{xvii} and \ion{Fe}{xxiv} lines for the calibration.
\ion{Fe}{xvii} and \ion{Fe}{xxiv} are blended in small flares, as discussed 
in \citealt{delzanna_etal:11_flare}.
In large flares (M-class), the 192.0~\AA\ line is normally saturated in 
the EIS observations. 
The  \ion{Fe}{xxiv} lines are strong during the impulsive phase of flares,
but often exhibit strong blue-shifted components which complicate the analysis.

\subsection{Silicon and sulfur lines }

\ion{Si}{x} lines are of particular importance for the calibration of the 
LW channel since they are among the strongest lines 
in quiet conditions, and there are three  good ratios to use.
We use the APAP atomic data of \cite{liang_etal:09_si_10}
and the QS observations.

The \ion{Si}{vii} 275.7 / 272.6~\AA\ is a good 
branching ratio. We use QS observations.
There are no \ion{Si}{vii} lines in the SW channel,
but \cite{kamio_mariska:12} and \cite{mariska:12_eis_cal} suggested the 
use of the \ion{Si}{vii} 275.3~\AA\ vs. \ion{Fe}{viii} 185.2~\AA\ ratio
to monitor the relative sensitivity of the LW/SW channels.
We will discuss this issue below.


 The \ion{S}{x} 257.1/ 264.2 ratio should  in principle be a good ratio 
(0.348 photons), but the weak 257.1~\AA\ line appears often blended,
especially in AR spectra, where at least the 
\ion{Fe}{xiii}  257.169~\AA\ line produces a contribution.
We use QS observations, where good agreement with the ground calibration
is found.
The \ion{S}{xi} 285.6 / 281.4~\AA\ is a good branching ratio in 
active region spectra.
The \ion{S}{viii} 198.55~\AA\ line is one of the few low 
temperature lines in the SW channel. $DEM$ modelling indicates that 
in quiet Sun regions this line is blended by about 30\% with 
\ion{Fe}{xi}.

\subsection{Data selection and processing}

The above discussion has made clear that a variety of 
targets were needed for the calibration.
There are surprisingly few good observations suitable for 
such calibration purposes. 
Very few EIS studies telemetred the full spectral range.
Most observations in the first few years were done with the 
1\arcsec slit and very low counts in the QS are present. 
Upon my suggestion, spectral atlases with the 2\arcsec slit 
(and long exposures)  have
been run since 2010 for monitoring purposes
(Atlas\_30, Atlas\_60, Atlas\_120).
They have been extremely useful for the present work and should be 
continued.

A selection of  QS data was chosen, by inspecting 
the spectra and avoiding  observations with strong hot lines.
This was relatively simple in the first three years of the mission,
but since then very few QS observations have been made. 
One further complication is that 
the typical QS disappears at temperatures above 1 MK 
\citep{delzanna_andretta:11} when solar activity increases.
A variety of EIS studies were selected.
A selection of active region (AR) observations was also  
chosen, to use the ratios involving 
the \ion{Fe}{xii}--\ion{Fe}{xvi} ions 
(avoiding saturated areas). A full list is given in the Appendix.

A few particular observations are worth mentioning.
To study the EIS sensitivity during the first month 
of routine operations (to limit instrument degradation), two dates were chosen.
A QS observation on 2006 Dec 23 was selected, with two regions,
one on-disk and one off-limb. 
An AR on-disk observation of hot core loops on 2006 Christmas day was also selected,
to build ratios of  the hotter lines.

For the flare lines, a suitable observation was searched but not found.
The search criteria were to find any EIS flare observation within the first 
8 months of at least a C-class flare and which contained the two 
\ion{Fe}{xvii}  and \ion{Fe}{xxiv} line ratios. 
So, the observations of 2007 Jun 2 discussed in \cite{delzanna:08_bflare} 
are considered here.

To increase the signal-to-noise, observations with long  exposures
have been selected. These have the drawback of increasing the incidence of cosmic rays.
The Solarsoft routine new\_spike, written by P.~Young, 
was used to automatically flag the cosmic ray hits as missing data.
The `hot' and `warm' pixels as listed within the EIS software 
were also flagged as missing data.
Each exposure was then visually inspected with custom-written software.
This was  necessary because occasionally the 
automatic routine does not detect cosmic rays, 
and also to note data dropouts or regions 
with high particle fluxes,  a common feature in  EIS observations.

The number of `hot' and `warm' pixels in the EIS CCD
has increased to a level such that since 2009 all observations need 
special processing. The missing data were replaced with 
interpolated data, obtained by successive interpolation along and across
 the slit direction, and visual inspection/manual corrections of each exposure.
This interpolation was necessary  before spatial averaging and line fitting.

 Various geometrical corrections  have been applied to the data
(see  \citealt{delzanna:09_fe_8} for details).
A `slant' in the spectra was found in \cite{delzanna_ishikawa:09}.
This results in a wavelength dependent offset along the slit
and has been corrected for by 
rotating each exposure and aligning the SW and LW channels. 
The offset of about 2\arcsec\  in the pointing of the SW and LW 
channels prior to 2008 Aug 24 was also taken into account.

Significant spatial averaging was needed to measure 
the weaker lines, given the low signal in most exposures.
Single-slit exposures were averaged along selected regions along the slit
(typically over more than 100\arcsec), while 
raster observations were first analysed to select the best regions
to obtain averaged spectra.  
QS regions were selected as those where no bright points or 
any strong hot (above 2 MK) emission was present. 
On the contrary, AR spectra were obtained from regions 
within the cores, where the \ion{Fe}{xiv} and \ion{Fe}{xvi} lines, needed for 
the calibration, were strong. Regions where the strongest lines are saturated
were avoided. 
Off-limb regions were selected by avoiding any particular structures.

Each of the resulting spectra was wavelength calibrated
(the EIS wavelength scale varies along the orbit).
All the EIS lines were fitted with Gaussian profiles
using the {\it cfit} package \citep{haugan97}, directly
on the raw spectra in data numbers (DN). 
One distinct feature of the EIS spectra is a variable
`bias'. 
 In particular, the SW channel shows an enhanced 
`background' in the center, which we believe to be 
mostly due to a pseudo-continuum of spectral lines.
The way the background is chosen affects the spectral 
fitting of the weak lines.
 Typical uncertainties in the lines are small,
of the order of a few percent for the strong lines.
Uncertainties were estimated by summing in quadrature the 
photon noise in the line and in the background, 
and by taking into account the spatial averaging. 
All the fits were checked visually.

From the the total number of counts in a line $N$ (DN),
the calibrated observed intensity $I_{\rm o}$ 
(phot cm$^{-2}$ s$^{-1}$ arcsec$^{-2}$)   is obtained by
(see EIS  software note No.2 for details)

$$ I_{\rm o} = {3.65  \; N \; \lambda \; G  \over 12398.5 \; \Omega \; E(\lambda) \; t } $$

where $\lambda$ it the wavelength in \AA, $t$ the exposure time in seconds, 
$\Omega$ is the solid angle subtended by each pixel in arcseconds square
(1 for the 1\arcsec\ slit, 2 for the  2\arcsec\ slit), 
12398.5 a conversion factor, 
$G$ is the gain of the CCDs, assumed to be 6.3,
3.65 the number of eV to produce an electron-hole in the 
CCD (silicon-based),
$E(\lambda)$ is the effective area, i.e. the area of the aperture 
multiplied by the various geometrical factors and 
the transmission coefficients of the optical
parts of the instrument (the  transmission of the filters, 
the mirror  reflectivity, the grating reflectivity, etc.).

The ground end-to-end calibration at RAL (UK) \citep{lang_etal:06}
 only provided measurements at one 
wavelength for the SW (205.9~\AA) and four wavelengths
for the LW (251.3, 256.3, 267.25, 283.4~\AA).
These measurements were compared to those predicted 
 by combining all the  efficiencies of the various optical components,
which were also measured in the laboratory. 
Significant discrepancies between the 
measured and predicted responsivities were reported by \cite{lang_etal:06},
both in terms of absolute values and in terms of relative values
within each channel.
The disagreement in the absolute values was not surprising, considering 
the large   uncertainty in  the quantum efficiency of the detectors.
Scaling the predicted values (by a 1.6 factor) however 
 still left a discrepancy of about 20\% between the SW and the 
LW channel, and about 50\% within the four measurements in the LW channel.
\cite{lang_etal:06} suggested the use of new effective areas, obtained by combining
 the  five calibration points with the predicted shapes of the responsivities.
An overall uncertainty of 22\% was estimated.
These effective areas  are available via {\it SolarSoft} 
(files EIS\_EffArea\_A.004 and EIS\_EffArea\_B.004).

\section{Results}

\subsection{SW and LW line radiances }

\begin{figure}[!htbp]
\centerline{\epsfig{file=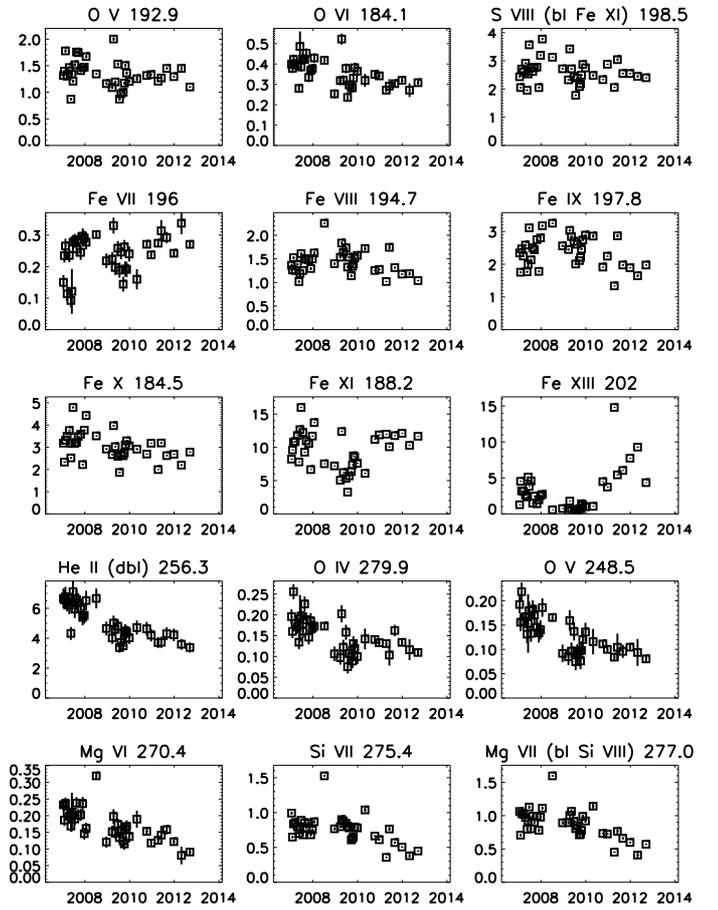, width=9.cm,angle=0 }}
  \caption{Averaged EIS count rates (DN/s per 1\arcsec) in the QS areas
as a function of time. The  \ion{He}{ii} 256.3~\AA\ line has been deblended.}
 \label{fig:quiet_int}
\end{figure}

\begin{figure}[!htbp]
\centerline{\epsfig{file=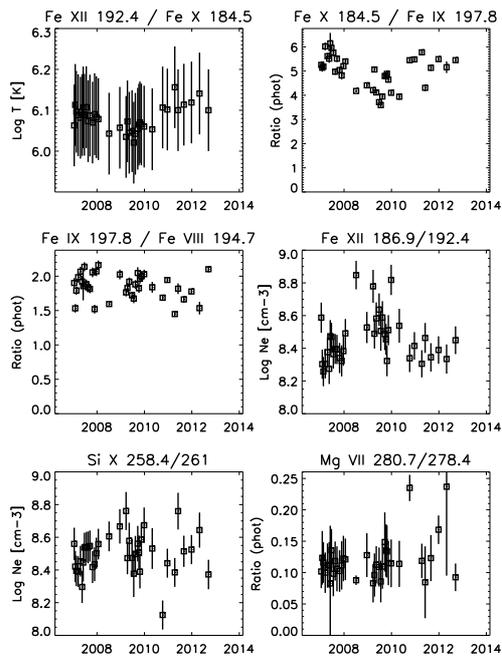, width=6.5cm,angle=0 }}
  \caption{Ratios of a few lines (photon units, using the ground calibration),
 with densities and temperatures obtained from them.}
 \label{fig:quiet_ratios_tene}
\end{figure}

We first present the averaged raw count rates in the QS areas in 
Fig.~\ref{fig:quiet_int}. The large scatter is mainly due to 
solar variability, however the cooler transition-region (hereafter TR) 
lines in the SW channel
do show a remarkable constancy over time, indicative of
no significant long-term degradation. The main line is the 
strong \ion{O}{v} self-blend, close to the peak of the SW effective
area, but other strong TR lines from \ion{S}{vi},  
\ion{Fe}{vii}, \ion{Fe}{viii}, \ion{Fe}{ix}, all show the same trends.
 Having said that, lines short-ward of 186~\AA\ 
(such as \ion{O}{vi} 184.1~\AA, shown in the figure)  do show a 
decrease, indicative of a small degradation at the 
shorter wavelengths. This is further discussed below.

The constancy of the TR line radiances and the 
direct EUNIS/EIS measurements of \cite{wang_etal:11}  suggest that 
the {\it absolute  SW central responsivity has not decreased significantly
over time}. This is assumed in the remainder of the paper.
Note that this result is in stark contrast to the conclusions 
reached by \cite{kamio_mariska:12} and \cite{mariska:12_eis_cal}.
We will return to this point in the conclusions.


All the cool lines in the LW channel do however show a decrease in their
QS radiances, as also shown in Fig.~\ref{fig:quiet_int}.
The \ion{He}{ii} 256.3~\AA\ line 
(which is actually a self-blend) is the strongest line in the channel
and shows a marked decrease,
by about a factor of two within the first two years. 
The  \ion{He}{ii} 256.3~\AA\ line is  blended with several
transitions on its long wavelength side, 
mainly due to \ion{Si}{x}, \ion{Fe}{x} \citep{delzanna_etal:04_fe_10}, 
\ion{Fe}{xii} \citep{delzanna_mason:05_fe_12}, and \ion{Fe}{xiii}
\citep{delzanna:11_fe_13}. 
Careful deblending of off-limb observations indicates the possible 
presence of further minor blending. 

Estimating the blends is not simple, as  the lines are density-sensitive. Fortunately,
the main contribution is from the \ion{Si}{x} 256.4~\AA\ line,  forming
a branching ratio with the \ion{Si}{x} 261~\AA\ line, which can be estimated
quite accurately. 
The broad profile of the \ion{He}{ii} blend was fitted with two Gaussians,
leaving all parameters to vary. The total intensity in the blend was then 
obtained by summing the two intensities. 
The contributions from the other ions was then subtracted,
using the measured (calibrated) intensities in the  \ion{Si}{x} 261,
\ion{Fe}{x} 257.2, \ion{Fe}{xii} 192.4, and \ion{Fe}{xiii} 252.0~\AA\ lines.
The result is that,  in on-disk QS observations, the 
\ion{He}{ii} dominates by contributing over 80\% to the observed
intensity. A similar result was found by \cite{mariska:12_eis_cal}, where 
however the \ion{He}{ii} 256.3~\AA\ line intensity was simply estimated by a 
double Gaussian fit.

As pointed out by \cite{kamio_mariska:12} and \cite{mariska:12_eis_cal},
 significant solar cycle changes are present in lines formed above 1 MK,
(such as  \ion{Fe}{xiii}), as expected 
\citep{delzanna_etal:10_cdscal, delzanna_andretta:11}.
Before  proceeding further, it is then important to estimate 
how the solar cycle variations can affect densities and 
temperatures in the QS observations considered here. 
This is shown in Fig.~\ref{fig:quiet_ratios_tene}.
The count rates were converted into calibrated radiances 
(photon units) using the ground calibration.

An estimate of temperature changes was obtained from the 
\ion{Fe}{xii} 192.4~\AA\ / \ion{Fe}{x} 184.5~\AA\ ratio, using CHIANTI v.6 ionization
equilibrium. Small changes within log $T$[K]=6--6.2 
are present, with decreasing temperatures toward the minimum on 2008,
and an increase. The \ion{Fe}{x} / \ion{Fe}{ix} ratios also show a similar
behaviour, which is simply due to the fact that lines formed above
1~MK increase their intensities.
The \ion{Fe}{ix} / \ion{Fe}{viii} ratios
 indicate very little changes in the upper QS TR temperature, instead.
Densities have also been measured from two among the best EIS ratios,
one from \ion{Fe}{xii} and one from \ion{Si}{x}, using lines 
close in wavelength, to reduce degradation effects.
The densities of the two ions agree very well (with the new atomic data 
for both ions), and show  little variability around log Ne [cm$^{-3}$]=8.4. 
The \ion{Mg}{vii} ratio
(see  \citealt{delzanna:09_fe_7} for details on blending issues) 
also indicates no significant changes in the  density of the 
QS upper transition region.

\def\baselinestretch{1.}

\begin{table*}[!htbp]
\caption{Quiet Sun historical irradiances and radiances.}
\begin{center} 
\footnotesize
\begin{tabular}{@{}lllllll@{}}

 \hline\hline \noalign{\smallskip}
                 & log $T$[K] & MH73 & H74  & M76  &  PEVE &  EUNIS07 \\
                   F10.7cm & & 177  & 123  & 102 &  69 &  67 \\
\hline \noalign{\smallskip}
\ion{He}{ii}           237 & & 0.44 & 0.57   & -    & 0.46 & - \\
\ion{He}{ii}           243 & & 1.1  & 0.95   & -    & 0.89 & -  \\
\ion{He}{ii}           304 & & -    &   72(bl, 2348)   & - & 63(bl, 2054) &  55 (1784) \\

\ion{He}{ii} (dbl)  256.3  &   & ?4.1 (?101) & 3.0 (74$^*$) & -   &  2.0 (49) &  - \\

\ion{O}{vi} 184.1        & 5.5 & 0.12 (3.0$^*$) &  -   & -   &  - &   - \\

\ion{Fe}{viii} 185.21    & 5.6& 0.44 (11) &  -   & -   &   0.35 (9) &  0.32 (8$^*$) \\ 

\ion{Si}{vii}  275.35     &5.75   & 0.28 (7) & 0.25 (6$^*$) &  -  &   0.35 (9) & -  \\ 

\ion{Mg}{vii} (bl) 278.40 &5.8&   0.34 (8) & 0.36 (9$^*$) &  -  &   0.33 (8) & - \\ 
\ion{Fe}{ix}   171.06     &5.85&   4.4 (109) & 4.2 (104$^*$) &  4.4 (109) & 7.0 (173$^*$) & -  \\ 

\ion{S}{viii} (bl) 198.53      &5.9  & 0.23 (6$^*$) &  -   & -   &   0.26 (6.4) &   - \\

\ion{Fe}{x}   174.53      &6.0  & 4.6 (114) & 4.1 (101) & 4.2 (104) &   3.7 (91) & 4.3 (107$^*$) \\
\ion{Fe}{x}   177.2       &6.0  & 2.6 (64)  & 2.2 (54)  & 2.6 (64)  &  2.3 (57) & 2.2 (55$^*$) \\
\ion{Fe}{x}   184.52      &6.0  & 1.0 (25)  & 1.2 (30)  & 1.0 (25)  &  1.2 (30) & 1.1 (27$^*$) \\
\ion{Fe}{x} (bl)  190.0   &6.0  & 0.53 (13) & 0.55 (14) & 0.57 (14) & 0.65(16) & 0.51 (12.5$^*$) \\
\noalign{\smallskip}

\ion{Fe}{xii}   193.50    &6.1  & 3.2 (80) & 3.1 (77) & 2.4 (59) & 1.0 (25) &  0.88 (21.7$^*$) \\
\noalign{\smallskip}\hline  
\end{tabular}
\normalsize
\tablefoot{The columns indicate the main ion and wavelength (\AA), the approximate 
formation temperature (log $T$[K]), then the MH73, H74, M76, PEVE irradiances 
(10$^8$ phot cm$^{-2}$ s$^{-1}$). The values in brackets indicate the estimated 
QS radiances (phot cm$^{-2}$ s$^{-1}$ arcsec$^{-2}$).
The last column gives the EUNIS07 measured QS radiances (in brackets), and
the QS irradiances estimated from them.
The 10.7 cm radio flux in 10$^{-22}$ W m$^{-2}$ Hz$^{-1}$ is also shown.
The values with an asterisk are those chosen as reference. 
 }
\end{center}
\label{tab:list_irr}
\end{table*}


\subsection{Calibrated radiances}

\begin{figure}[!htbp]
\centerline{\epsfig{file=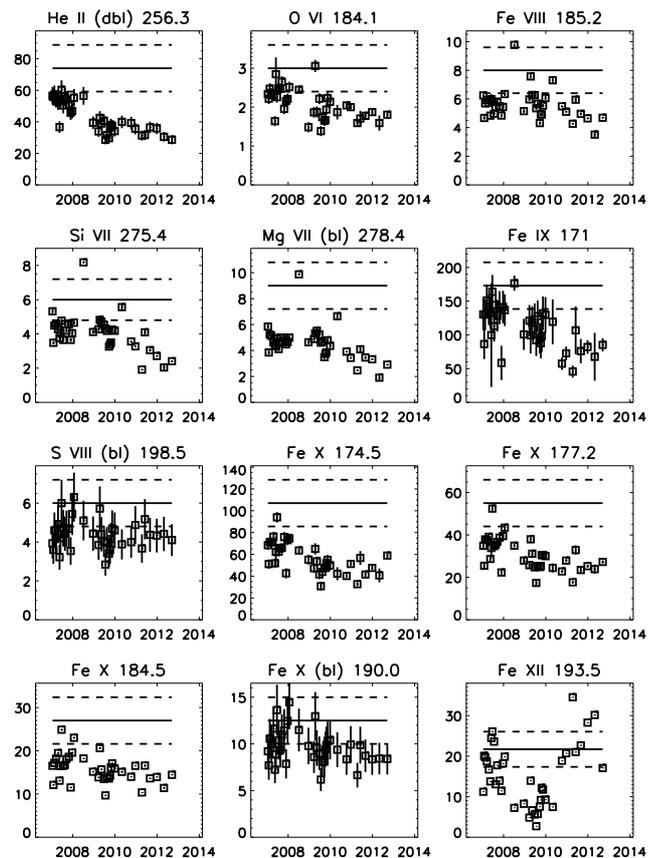, width=8.5cm,angle=0 }}
  \caption{Averaged EIS QS radiances (phot cm$^{-2}$ s$^{-1}$ arcsec$^{-2}$)
obtained with the ground calibration. 
The lines indicate the QS radiances listed in Table~\ref{tab:list_irr}
(the dashed lines with a $\pm$20\%). 
 The  \ion{He}{ii} 256.3~\AA\ line has been deblended.}
 \label{fig:quiet_rad_std}
\end{figure}

\begin{figure*}[!htbp]
\centerline{\epsfig{file=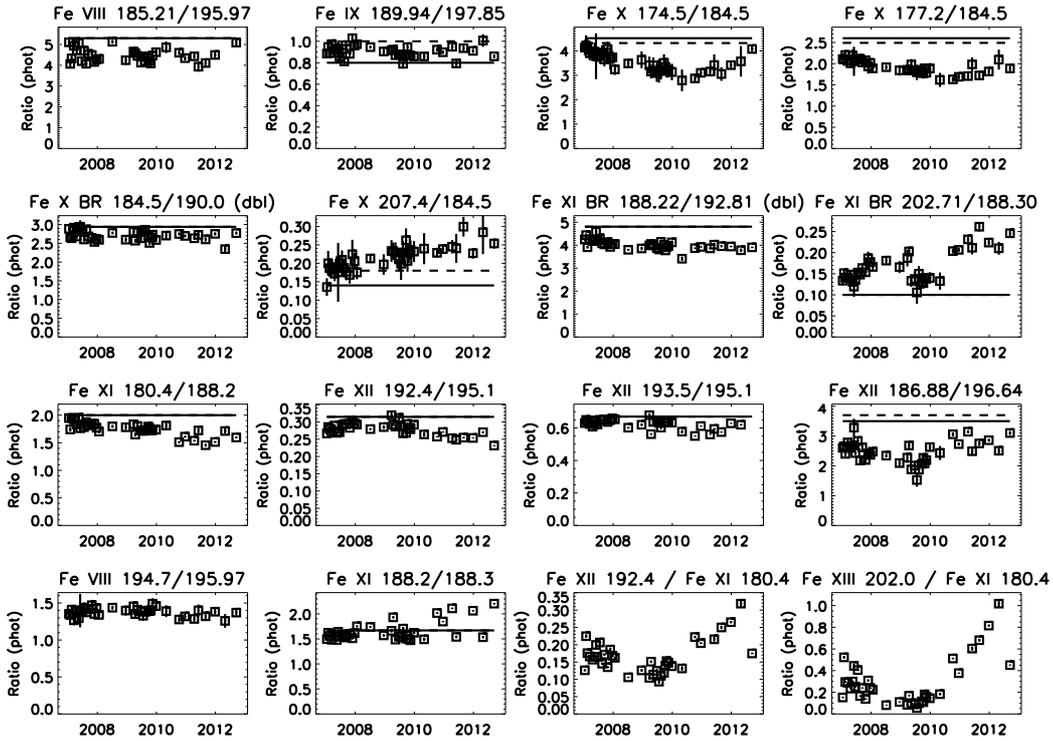, width=10.cm,angle=90 }}
  \caption{Ratios of a few SW QS line radiances (photon units) using the 
ground calibration. Bars indicate 
the predicted  values. }
 \label{fig:quiet_ratios_sw}
\end{figure*}

One question naturally arises: how do the calibrated radiances
compare with previously well-calibrated measurements for the stronger lines?
 As discussed in \cite{delzanna_etal:10_cdscal,delzanna_andretta:11} 
lines formed in the transition region up to 1~MK have been found to have irradiances
that are not significantly affected by solar activity (presence of active regions),
and are also expected to have radiances in quiet Sun regions that are approximately
constant. So although solar variability is always present, by spatially averaging 
over large regions one should obtain EIS radiances in agreement 
with other well-calibrated measurements. 

 As discussed in \cite{delzanna_andretta:11}, the best 
calibrated irradiances in the EUV are those obtained by 
\cite{malinovsky_heroux:1973} [MH73], \cite{heroux_etal:1974} [H74],
together with \cite{manson:76} [M76], and those we obtained from PEVE, 
the prototype of the Solar Dynamics Observatory (SDO) 
 Extreme ultraviolet Variability Experiment (EVE), flown on  2008 April 14
during the deep solar minimum \citep{woods_etal:09,chamberlin_etal:09}. 
These values are shown in  Table~\ref{tab:list_irr}. 
The \ion{He}{ii}  256.3~\AA\ line has been deblended from the strong
\ion{Si}{x} contribution using the other \ion{Si}{x} lines and theory,
for both MH73 and H74 data, although in principle MH73 lists 
5.2$\times$10$^8$ phot cm$^{-2}$ s$^{-1}$ as already deblended. 
The fact that good agreement exists between MH73 and H74 for the 
other \ion{He}{ii}  237 and 243~\AA\ lines (see Table~\ref{tab:list_irr})
suggests a mistake in 
MH73's list. The PEVE measurement of the  256.3~\AA\ line is quite uncertain
given the low resolution of the instrument.

It is well known that it is possible to obtain from the irradiances an estimate of the 
QS radiances at disk center, once a correction factor for the limb-brightening  is known. 
It is well known that  helium lines do not show any limb-brightening, while 
all the TR lines do. We have recently measured \citep{andretta_delzanna:2012}
these limb-brightenings using the SOHO/CDS radiances for a range of ions,
and found corrections factors close to 1.4 for all lines formed in the lower and 
upper TR, up to 1~MK. We also estimated the correction due the off-limb contribution 
  \citep{delzanna_etal:10_cdscal}, but this is negligible (a few percent 
maximum) for these lines. We have therefore applied a 1.4 correction factor to obtain 
estimated QS radiances from the irradiance measurements.
These QS radiances are also shown in Table~\ref{tab:list_irr} in brackets.
In terms of actual QS radiances, very few observations at the EIS 
wavelengths exist. There are the SOHO GIS observations 
(for example \citealt{delzanna_thesis99,andretta_etal:03}), 
although the GIS lines were 
calibrated with the line ratio technique \citep{delzanna01_cdscal}.
It is interesting to note that good agreement is found between these
GIS measurements and the EIS ones. 
The SERTS-95 rocket flight also measured  QS radiances for a few of 
the brightest lines, but older CHIANTI atomic data were used for its 
calibration using again the line ratio technique \citep{brosius_etal:98a}.
Then there are the recent EUNIS 2007 rocket flight measurements of 
\cite{wang_etal:11}, which were instead calibrated on the ground.
These values are reported in Table~\ref{tab:list_irr}.
Overall, there is good agreement between the  EUNIS 2007 radiances
and those obtained from historical irradiances. 
The values with a star in Table~\ref{tab:list_irr} are those adopted 
here for comparison with the EIS radiances.
There is also good agreement among the various irradiances, with the notable 
exception of the strong \ion{Fe}{ix} 171.06~\AA\ 
and the deblended \ion{He}{ii}  256.3~\AA\  lines. The variations
in the  \ion{He}{ii} are most likely caused by the different solar 
activity  \citep{delzanna_andretta:11}, but the \ion{Fe}{ix} is a puzzle,
with PEVE providing higher irradiances.
Various checks have been made on the EUV lines observed by PEVE,
and no problems in the calibration have been found. 
So the PEVE measurement has been adopted, considering the thorough
calibration work that was done before and after that flight.

The EIS QS radiances for a selection of TR lines for which 
we have previous measurements are shown in Fig.~\ref{fig:quiet_rad_std}.
They were obtained with the ground calibration, and clearly show 
values that are  lower (by 20--30 \%) than previous data for most wavelengths, 
even at the beginning of the mission, when the Sun was very quiet. 
We recall that the EUNIS 2007  measurements were done during the 
very low minimum, so these radiances should represent the lowest
QS values.
The only explanation for the discrepancies is that 
at the wavelengths displayed the EIS sensitivity in 2007 had 
already decreased compared to the pre-flight values,
as already suggested by \cite{wang_etal:11}.
We will return to this issue below.

\subsection{SW and LW line  ratios}

\begin{figure}[!htbp]
\centerline{\epsfig{file=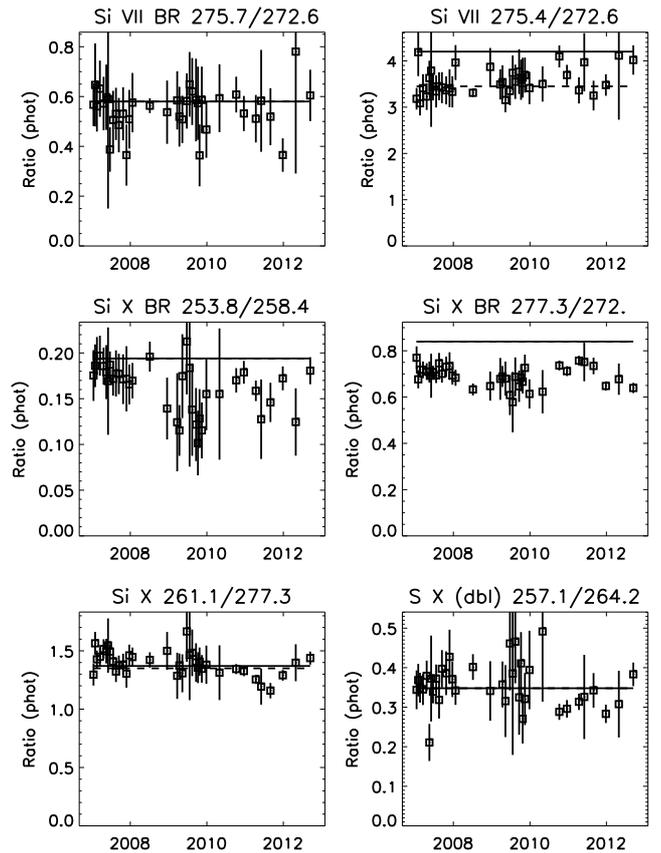, width=8.5cm,angle=0 }}
  \caption{Ratios of a few LW QS  line radiances (photon units) using the 
ground calibration. Bars indicate 
the predicted  values. }
 \label{fig:quiet_ratios_lw}
\end{figure}

\begin{figure}[!htbp]
\centerline{\epsfig{file=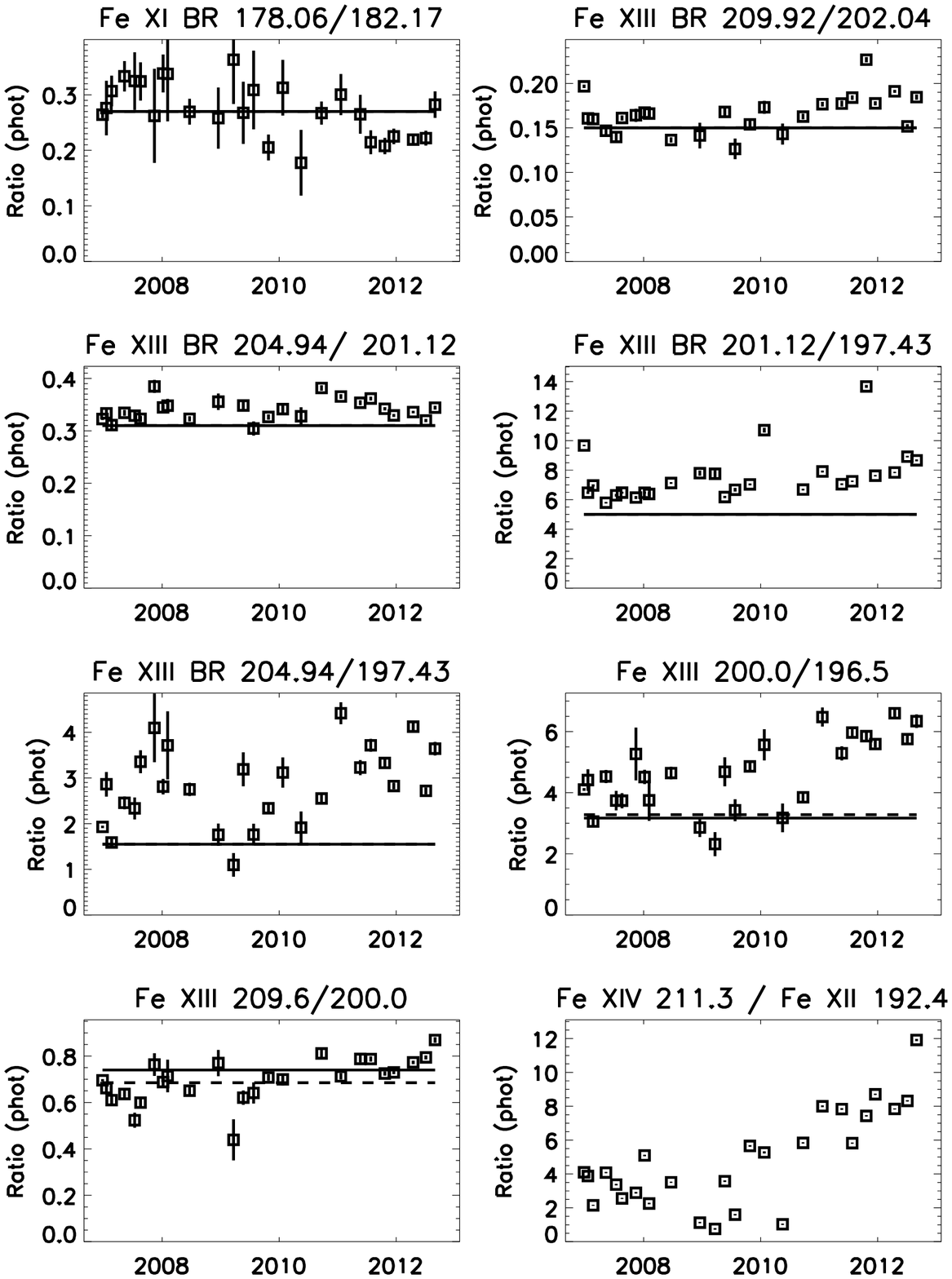, width=8.5cm,angle=0}}
\centerline{\epsfig{file=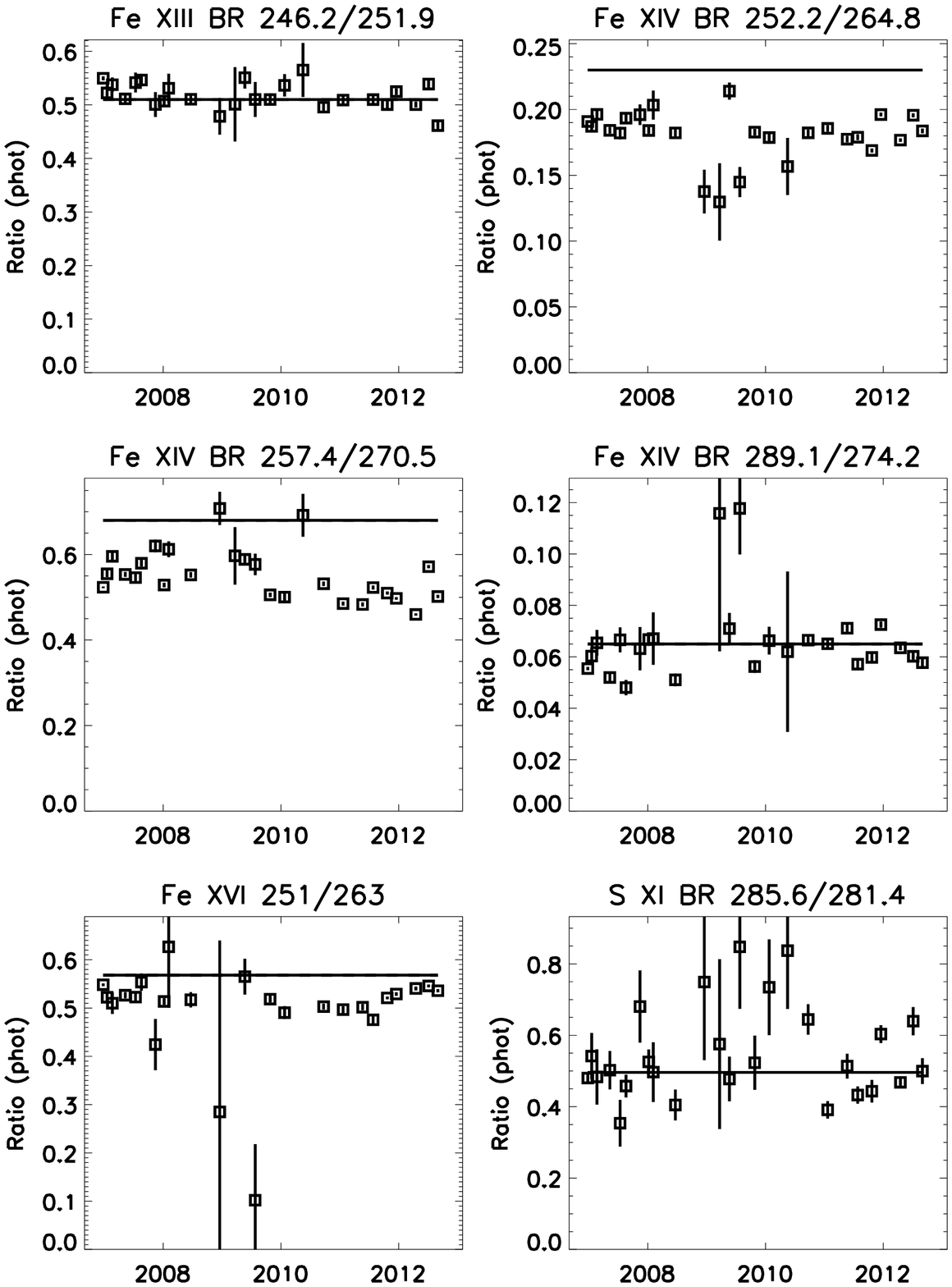, width=8.5cm,angle=0 }}
  \caption{Ratios of a few  AR line radiances (photon units)
in the SW and LW channels, obtained with  the ground calibration. Bars indicate 
the predicted  values. }
 \label{fig:active_ratios}
\end{figure}

Figs.~\ref{fig:quiet_ratios_sw},\ref{fig:quiet_ratios_lw} 
show a selection of line ratios within each  SW and LW channels, again 
using the ground calibration.
 Bars indicate the predicted  values, for two different densities,
log Ne [cm$^{-3}$]=8 (full lines) and log Ne  [cm$^{-3}$] =9 (dashed lines).
The line ratios from AR observations (both on-disk and off-limb cases 
have been considered) are shown in Fig.~\ref{fig:active_ratios}.
AR spectra offer more  branching ratios and  better S/N.

These line ratios hold a lot of information. 
Firstly, among those that are expected to be constant in time,
most of them do not show significant trends. This  is a strong
indication that for most wavelengths the relative sensitivities within each SW and 
LW channel have not changed over time. 
This supports the analysis performed in the following section,
where the shapes of the effective areas are obtained for the 
beginning of the mission from observations during the first two years.

Secondly, the small scatter in many of them confirms the correct identification
and absence of blending (or correct deblending applied whenever the case). 
Thirdly, the ratios are generally in  good agreement with theory, 
which gives confidence in both
the atomic data and the accuracy of the EIS effective area curves
within its stated uncertainty (22\%) for many wavelengths. 
 The ratios indicate that the shorter wavelengths of both channels 
had degraded, since they are consistently lower than expected.

Clearly, some ratios show puzzling departures.
A few SW ratios such as the \ion{Fe}{x} 174.5/184.5~\AA,
the \ion{Fe}{x} 177.2/184.5~\AA\ and 
the  \ion{Fe}{xi} 180.4/188.2~\AA\  in Fig.~\ref{fig:quiet_ratios_sw}
show a decrease suggesting that a further 
degradation occurred in the   170--184~\AA\ region at least in the first few years.
Other ratios we looked at have significant departures beginning in 2010,
such as the branching ratio \ion{Fe}{xi} 202.71/188.3~\AA\ 
shown in Fig.~\ref{fig:quiet_ratios_sw}.
In this case, either the  \ion{Fe}{xi} 202.71~\AA\ becomes 
blended or the 188.3~\AA\ decreases its intensity (or both).
The same figure shows that the \ion{Fe}{xi} 188.2/188.3~\AA\ is relatively
constant (but with large scatter), so it is possible that the 
 \ion{Fe}{xi} 202.71~\AA\ lines becomes blended with a higher
temperature line from e.g.  \ion{Fe}{xii} or  \ion{Fe}{xiii}
(\ion{Fe}{xiv} lines are excluded because they are comparatively
weak in the QS data). 
Fig.~\ref{fig:quiet_ratios_sw} shows in the bottom right corner
that  \ion{Fe}{xii}/\ion{Fe}{xi} and  \ion{Fe}{xiii}/\ion{Fe}{xi} ratios 
increase significantly after 2010, due to 
 the QS solar corona becoming affected by the solar cycle.

As in the QS case, some AR ratios that are expected to be constant 
are not. One example is the \ion{Fe}{xiii}  200.0/196.5~\AA\ ratio,
shown in Fig.~\ref{fig:quiet_ratios_lw}. The increase could be 
due to  a line from an ion with higher formation temperature such as 
\ion{Fe}{xiv}. In fact, the same figure shows that the 
\ion{Fe}{xiv} 211~\AA\ /  \ion{Fe}{xii} 192.4~\AA\ increases
after 2010. These two lines are strong and not significantly blended,
which means that on average the \ion{Fe}{xiv}emission tends
to increase compared to the \ion{Fe}{xii} in the active regions cores.

\subsection{SW/LW ratios and long-term correction}

\begin{figure}[!htbp]
\centerline{\epsfig{file=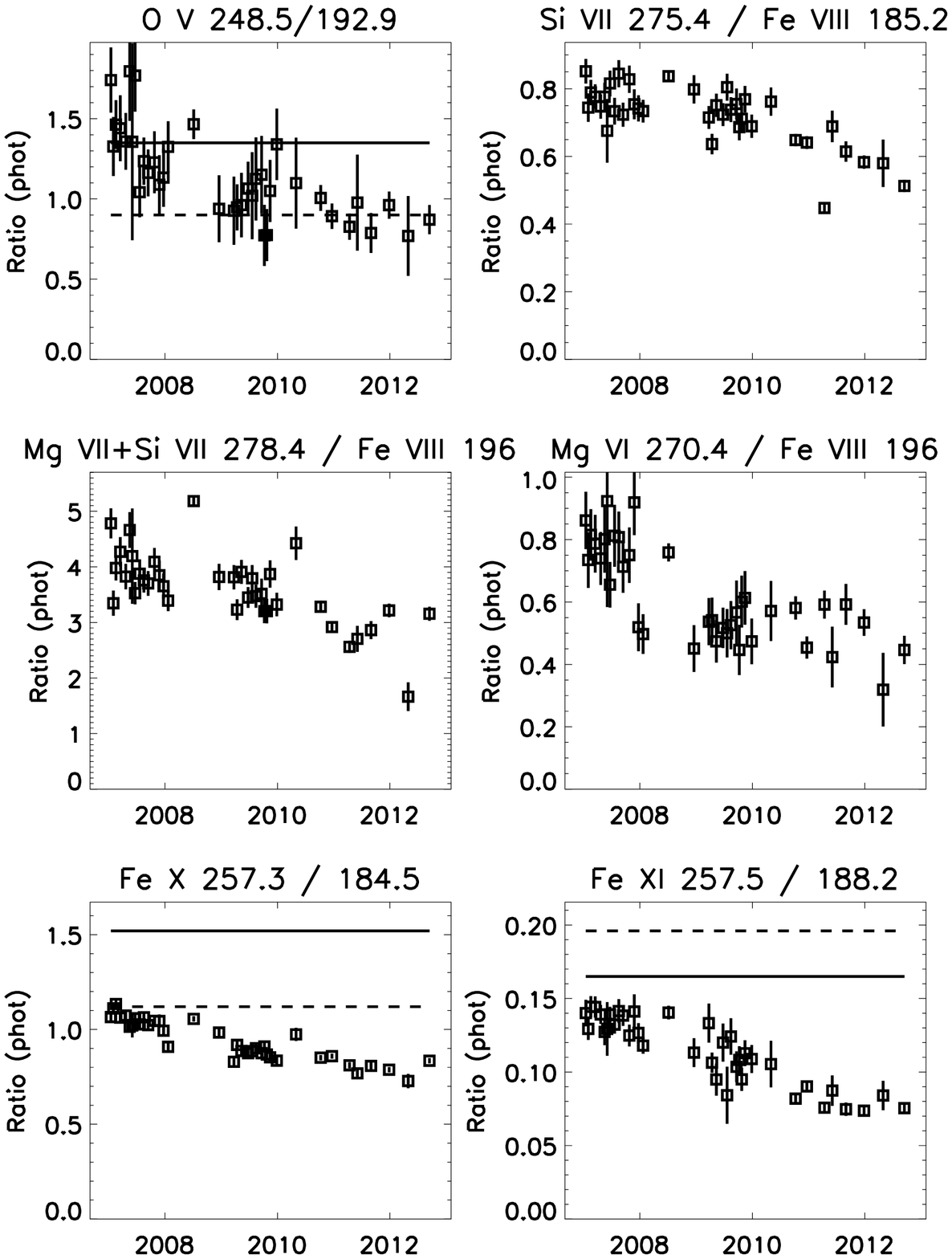, width=7.cm,angle=0 }}
 \centerline{\epsfig{file=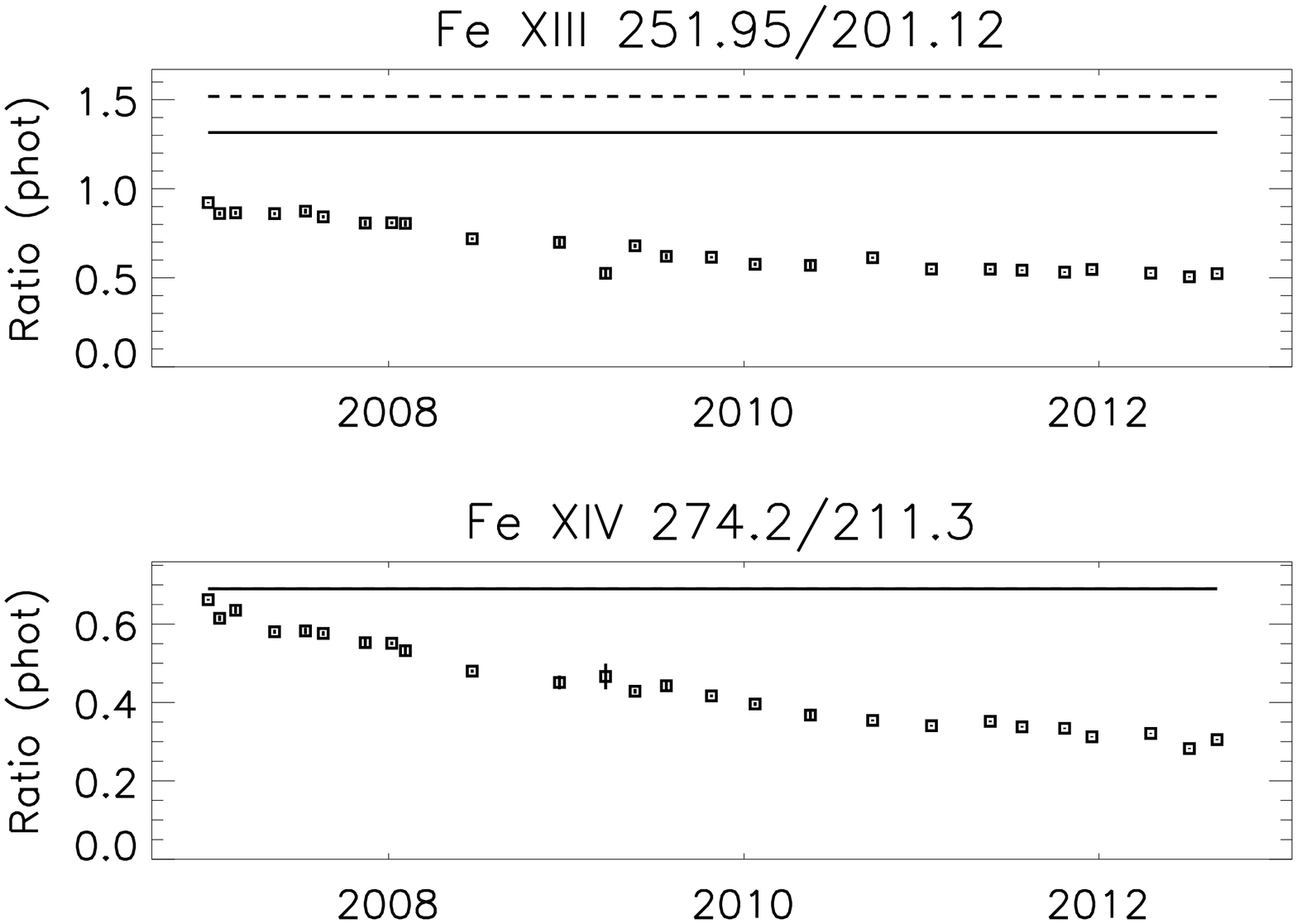, width=6.cm,angle=90 }}
 \caption{Ratios of a few LW vs. SW line radiances (photon units) using the 
ground calibration. Bars indicate 
the predicted  values. The top six plots shows the ratios from the QS observations,
the lower ones are from the AR observations. }
 \label{fig:ratios_sw_lw}
\end{figure}
 
We now examine the cross-calibration between the SW and LW channels.
A selection of ratios in  lines in the LW and SW channels 
 is shown in Fig.~\ref{fig:ratios_sw_lw}. The line intensities were
calculated using the ground calibration.
All the ratios, unlike the previous ones, show  decreases over time,
more marked in the first two years of the mission (a factor of two).
The  \ion{Fe}{xiv} 274.2 / 211.3~\AA, \ion{Fe}{xiii} 251.9 / 201.1~\AA, 
and \ion{Fe}{xi} 257.5 / 188.2~\AA\ ratios show a similar trend,
within a relative 20\%. This is a clear indication of a decrease in 
the LW/SW relative responsivities, as already seen with Fig.1. 
The \ion{Fe}{xiii}  201.1 and 211.3~\AA\ lines have been deblended. 
The \ion{Fe}{xiv} ratio is only slightly 
temperature sensitive, with an expected variation of less than 
10\%.

The ratios  involving lines in the 250--260~\AA\ 
region  show observed LW/SW  ratios lower than expected 
    (as already discussed previously), even at the start of the mission.
However,  the  \ion{Fe}{xiv} 274.2 / 211.3~\AA\ ratio is close,
at the start of the mission, to the expected value. 
This is another clear indication that the ground calibration 
should be revised, even for the start of the mission. This is 
discussed in the next section.

The \ion{Fe}{x} ratio shown in  Fig.~\ref{fig:ratios_sw_lw}
has some temperature and density dependence, however the expected variations 
based on the measured values are small (less than 10\%), and cannot 
explain the differences with the \ion{Fe}{xi} ratio. 
One explanation could be a  decrease in the sensitivity at  
 184.5~\AA. This would be in line with the 40\% decrease of the 
\ion{O}{vi} 184.1~\AA\ radiance, but not with the constancy of the 
184.5/190~\AA\  (and other) ratios.

The \ion{O}{v} ratio is not very reliable because the 248.5~\AA\ 
is very weak and because the ratio varies considerably with density.
%
%
The \ion{Si}{vii} 275.3~\AA\ vs. \ion{Fe}{viii} 185.2~\AA\ 
ratio shows a  slightly smaller decrease in time, compared to the 
other ratios, especially when the Sun becomes active.
This ratio was used by \cite{kamio_mariska:12} and \cite{mariska:12_eis_cal} to argue
that both LW and LW channels had the same decrease in sensitivity. 
Other similar ratios such as the 278.4/196~\AA\ show a similar behaviour.
On the other hand,  \ion{Mg}{vi} vs.  \ion{Fe}{viii} ratios are 
different. As we have seen, the the density and 
temperature of the QS upper transition region does not seem to change much, 
so differences in the various ratios should not be expected. 
The situation is actually  quite complex as discussed later.

\subsection{An estimate of the in-flight SW, LW effective areas}

\begin{figure}[!htbp]
\centerline{\epsfig{file=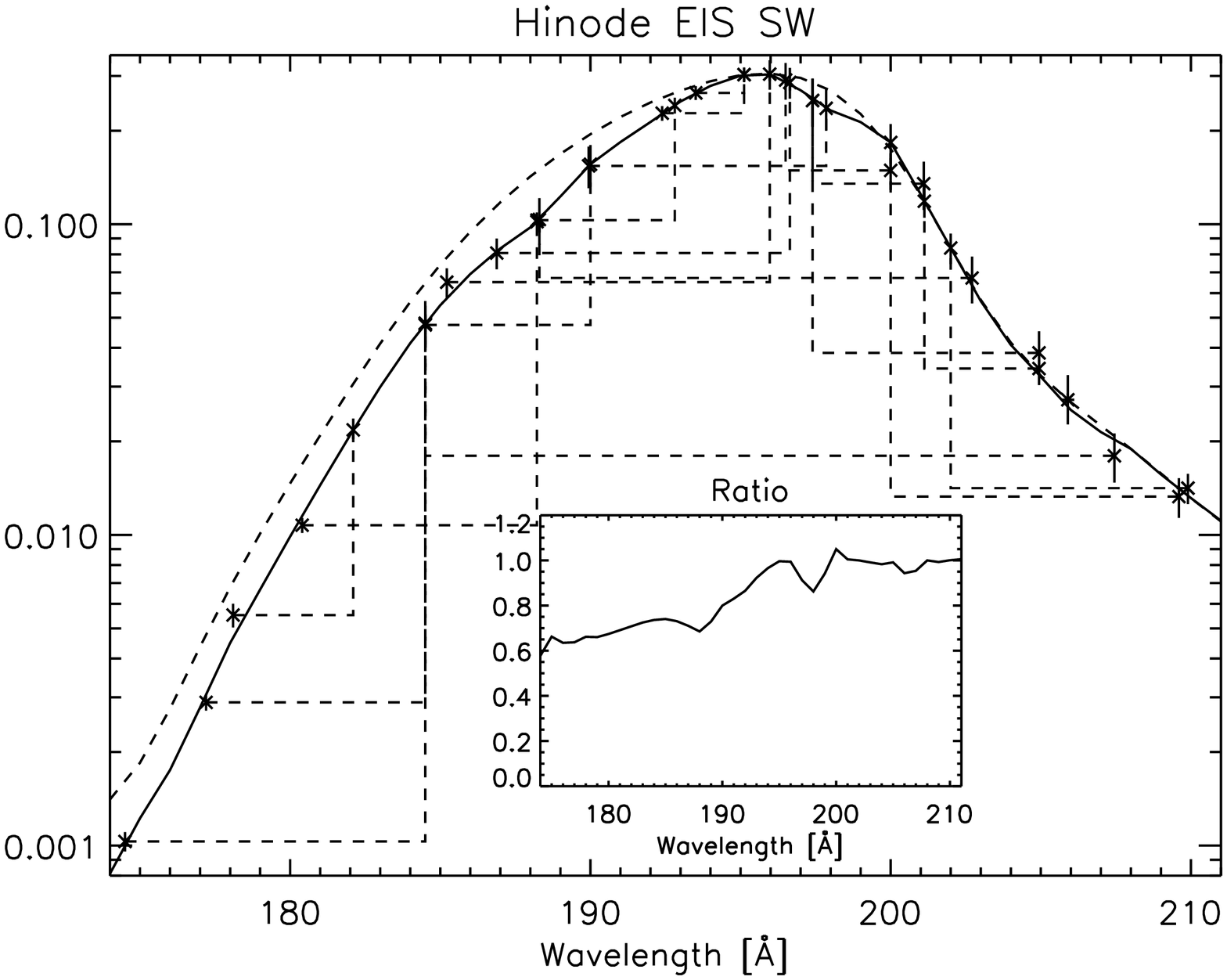, width=7.cm,angle=90}}
\centerline{\epsfig{file=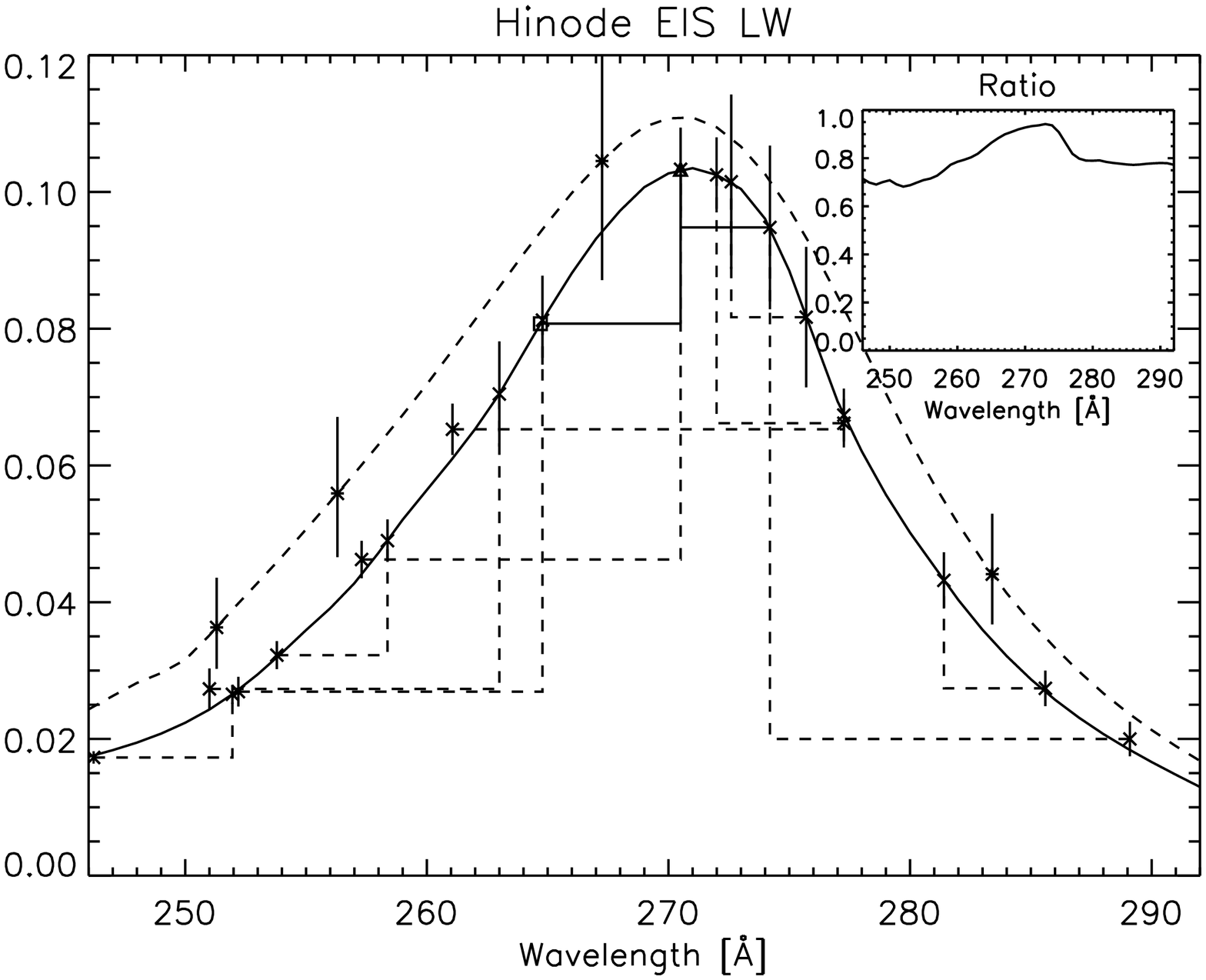, width=7.cm,angle=90}}
\centerline{\epsfig{file=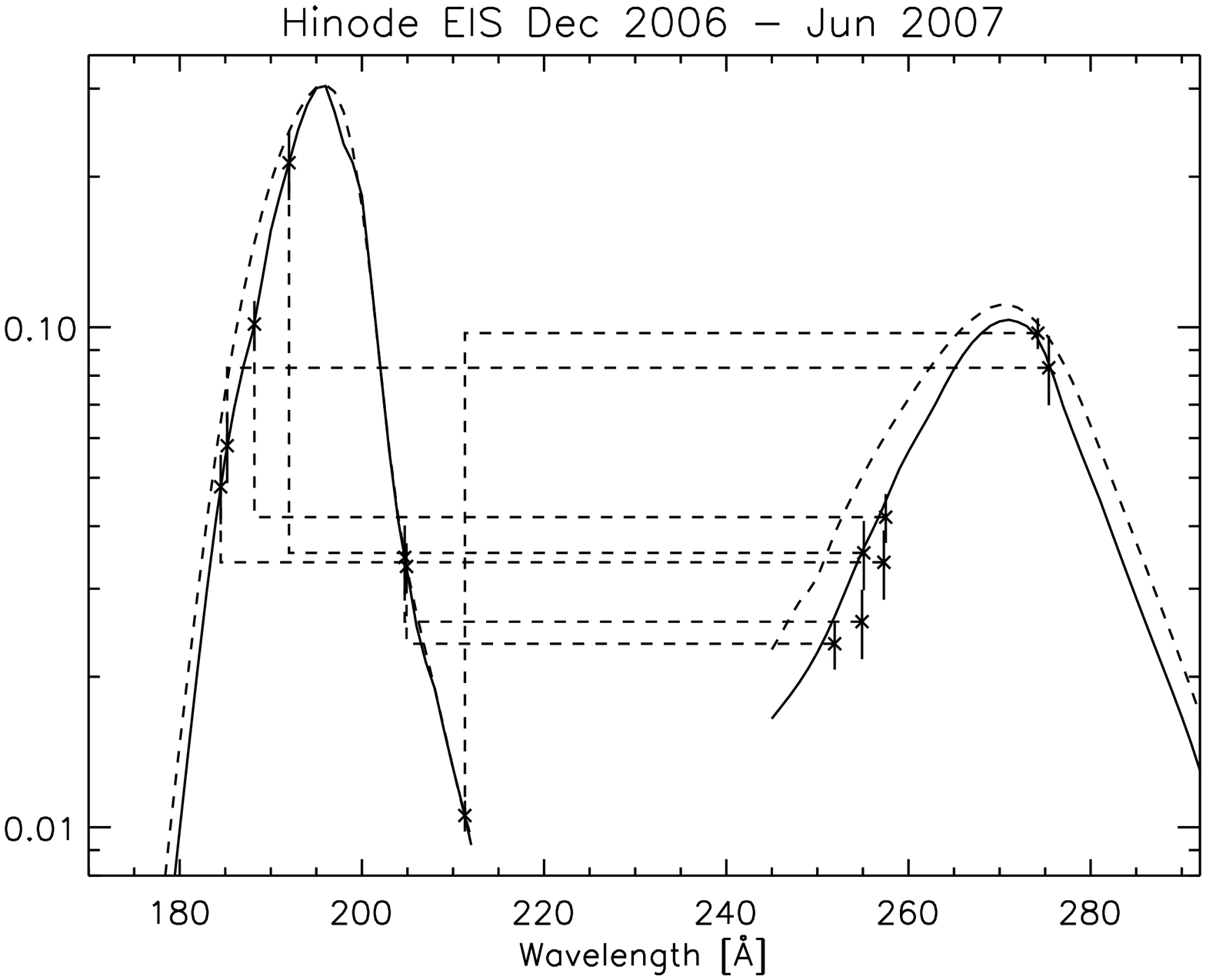, width=7.cm,angle=90}}
\caption{Effective areas (cm$^2$) for the EIS channels obtained from the 
 first two years of QS and AR observations analysed here, 
 and the 2007 Jun 2 flare observation \citep{delzanna:08_bflare}. The full lines indicate 
the proposed values, while the dashed ones indicate those
from the ground calibration. The two plots above show the 
line ratios used for the SW and LW channels separately, while the bottom
plot shows the cross-channel line ratios. 
The asterisks show the five measurements of the ground calibration.
The insets show the ratios
between the measured values and those of the ground calibration.
}
 \label{fig:eff_dec_2006} 
\end{figure}

Each observed and theoretical 
line ratio considered produces a constraint to the relative 
values of the effective areas. 
Considering two lines with theoretical intensities  $I_1$
and $I_2$, the ratio of the effective areas is obtained 
directly from the observed counts: 
$R_{\rm eff} (\lambda_1,\lambda_2) = E_1 / E_2 = I_2 \; N_1 \; \lambda_1 / (I_1 \; N_2 \; \lambda_2)$.
Table~\ref{tab:list_lines} lists the measured count rates (DN/s) in the pairs
of lines, the observed ratio and the derived  ratio of the effective areas $R_{\rm eff}$,
compared to that from the ground calibration.

The relative constancy  of the various  ratios for at least the 
first two years of the mission  means that it is possible to obtain one
set of in-flight effective areas that are valid for that period.
For each ratio listed in  Table~\ref{tab:list_lines}, we have 
considered the average over this period. 
For the ratios involving lower temperature lines we have considered
the QS values, while for the hotter ones the AR ones as described
previously. The average values are shown  in Table~\ref{tab:list_lines}.
The same table also shows the average count rates in each line
forming the ratio. 
Considering that the exposure time was 90 seconds,
the numbers in the table show that the total counts in most selected 
lines were high, a few hundreds. Hence,  single measurements of most lines 
have a small uncertainty. 
 We have adopted as uncertainty in the 
ratios one standard deviation from the average. These uncertainties
are also listed in Table~\ref{tab:list_lines}.
Finally, for the uncertainties in the 
 ratios of the effective areas $R_{\rm eff}$ we have summed in quadrature
the uncertainties in the averaged observed ratios, and the 
estimated theoretical uncertainties of the ratios, which are also 
listed in Table~\ref{tab:list_lines}.

We have done the procedure independently for the SW and LW channels. 
We started with two  splines at a selection of nodes 
and with values such that the curves were identical to the ground 
calibration curves. We then adjusted the spline node values 
to satisfy the observed ratios of the effective areas. 
The final effective areas are shown in  Fig.~\ref{fig:eff_dec_2006}
in the two top plots.
Each line ratio used is linked with dashed lines, and the 
combined uncertainties for each ratio overplotted.
The values of the spline nodes are listed in the Appendix.

The SW effective areas were scaled by assuming that the 
values near 195~\AA\ (the peak) are the same as those of the ground
calibration. 
The LW effective areas were scaled as described in the 
following section.
The five ground calibration measurements
(one for the SW and four for the LW) are also shown, with their
20\% uncertainty.

The results for the SW channel are shown in logarithmic scale.
Overall, good agreement with the ground calibration (dashed line) is found,
although significant departures in the shorter wavelengths of both 
channels is present.  This result is puzzling, but 
there is a significant consistency in the 
various ratios, and in all the subsequent analysis 
that has been done on several datasets.
Note that  the discrepancies are not within the estimated 
uncertainties. We will return to this issue in the conclusions.

\subsection{An estimate of the relative LW/SW effective areas}

The LW/SW relative values in  Fig.~\ref{fig:eff_dec_2006} (bottom plot) were 
obtained from a combination of measurements. We have used the 
off-limb QS of 2006 Dec 23 for the lower-temperature lines, 
the on-disk AR spectra of 2006 Dec 25 for the 
 \ion{Fe}{xiii} and \ion{Fe}{xiv} line ratios, and the 
 2007 Jun 2 flare for the \ion{Fe}{xvii}  and \ion{Fe}{xxiv} line ratios.
The measurements of the flare lines 
(listed in \citealt{delzanna:08_bflare}) are uncertain because of 
the offset of about 2\arcsec\  in the pointing of the SW and LW 
channels mentioned previously 
  and the dynamic nature of flares.
However, the flare was very small and the variations to be expected 
from the non-simultaneity of the observations cannot explain the 
 discrepancies, which are also found in every later observation
after 2008 Aug 24, as mentioned in the introduction.

In combining the 2006 Dec and 2007 June datasets we neglect 
the variation in the relative LW/SW calibration, which is however
small as described below.
The LW effective areas are considerably lower than measured
on the ground. As we have seen, they continued to decrease over time.

A detailed emission measure analysis was performed on the off-limb
quiet Sun observation of  2006 Dec 23. Excellent agreement between all
the SW and LW lines was found, further confirming the new calibration.
The results of this analysis are not presented here
since they  would require a lengthy discussion of all 
the blends and the lines. 
A similar line ratio study was performed with  the 2007 Aug 19 
AR observations discussed in detail in \cite{delzanna:12_atlas}.
The new effective areas remove all the main  discrepancies.

\subsection{The long-term correction}

We recall that the ratios shown in Fig.~\ref{fig:ratios_sw_lw} show similar
trends, i.e. a marked decrease in the LW/SW ratios
during  the first two years of the mission, when the Sun was very quiet,
hence no significant changes would be expected.
To provide an estimate for the long-term drop in sensitivity, 
we have considered the  \ion{Fe}{xiv},  \ion{Fe}{xiii}, and  \ion{Fe}{xi} 
ratios. Each of them has its pros and cons. 
AR spectra offer better signal, but a large number of unidentified
lines appear there (see the spectral atlas in \citealt{delzanna:12_atlas}),
 so it is possible that additional blending occurs 
in the lines. Additionally, AR spectra are normally very inhomogeneous, with large
variations in temperatures and densities which, even if slightly,
could affect some of the ratios. 
The Fe XIV 211 and 274 ratio has the problem that 
the 211 is just at the edge of the SW channel, and the 274 is blended
with a Si VII line, which can contribute up to 20\% or so.
This latter line is density sensitive and has additional uncertainty 
in its estimation.
The QS spectra have less signal, and show larger scatter.

\begin{figure}[!htbp]
\centerline{\epsfig{file=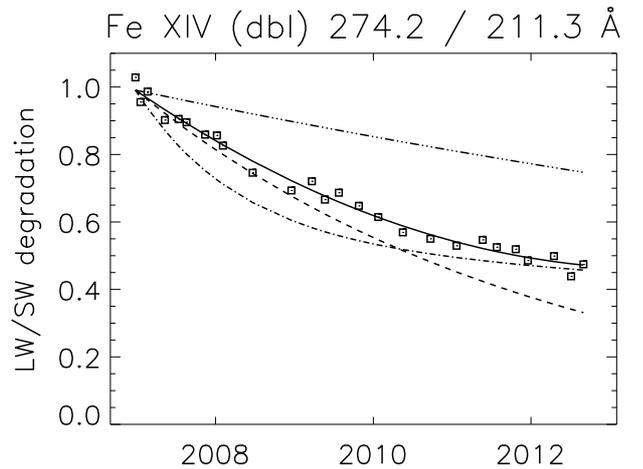, width=7.cm,angle=90 }}
  \caption{Observed  \ion{Fe}{xiv} 274.2 / 211.3~\AA\ ratio, used to 
infer the  long-term correction for the decrease in 
sensitivity of the LW channel. The full line is the fit, while the 
other lines are previous suggestions (see text). }
 \label{fig:lt_correction_linear}
\end{figure}

To provide an estimate for the long-term drop in sensitivity, 
we show  the results from the   \ion{Fe}{xiv} 274.2 / 211.3~\AA\ ratio
in Fig.~\ref{fig:lt_correction_linear}.
Similar results are obtained from the \ion{Fe}{xi} and \ion{Fe}{xiii}
ratios. Note that this agreement is there only if the recent atomic data
for \ion{Fe}{xi} are used, and careful deblending of the  \ion{Fe}{xiii} 
and  \ion{Fe}{xiv} is done.

The trend in the  \ion{Fe}{xiv} 274.2 / 211.3~\AA\ ratio is
easily measured, but the absolute value of the correction 
depends on the radiometric calibration and the predicted value for the 
ratio. We have adopted the present radiometric calibration described in 
the previous section, and recall that we assumed a predicted value of 0.69 for this ratio.
We note that the ground calibration provides similar values.
The ratio points have been fitted with a polynomial curve, also shown in 
 Fig.~\ref{fig:lt_correction_linear}. This curve is assumed as our long-term 
correction for the degradation of the LW channel (details are given in the Appendix).

The three other curves in  Fig.~\ref{fig:lt_correction_linear}
show for comparison the (normalised to the first point) exponential decays
as available within the EIS software (dashed line),  
as suggested by \cite{kamio_mariska:12} (dot-dashed line), 
and by \cite{mariska:12_eis_cal} (triple dot-dashed line).
We recall that both \cite{kamio_mariska:12} and 
\cite{mariska:12_eis_cal}  assumed the same decay rate for both SW and LW. For the
purposes of this plot these results have been interpreted as only applying to
the LW channel.
Note that there are clear differences in  the behaviour of the four corrections.

As a {\it first-order calibration} we assume in the remainder that:
a) the shapes of the SW and LW effective areas are as shown in 
Fig.~\ref{fig:eff_dec_2006} and do not change over time;
b) the SW sensitivity does not change over time;
c) the LW effective areas decrease  over time as shown 
by the fit in Fig.~\ref{fig:lt_correction_linear}.
The parameters for the polynomial fit
are given in Appendix B.


\subsection{The first-order calibration applied to a few cases}

\begin{figure}[!htbp]
\centerline{\epsfig{file=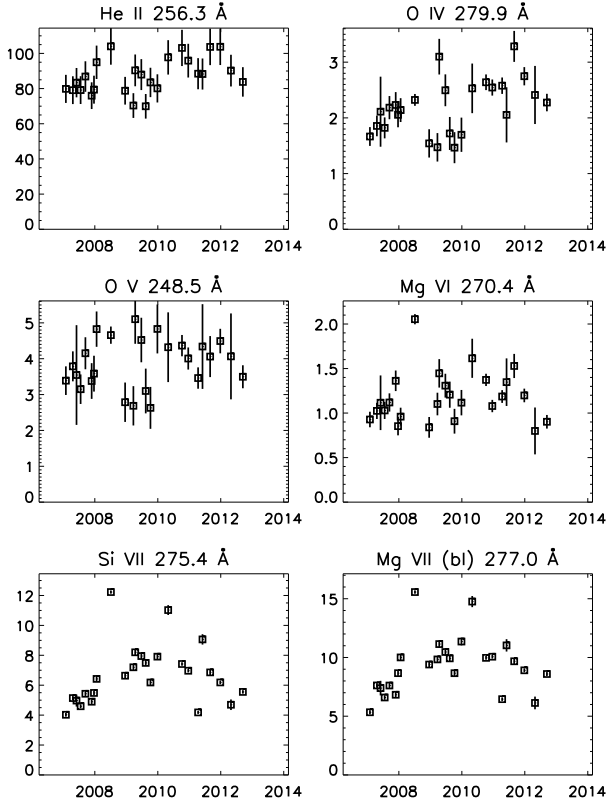, width=8.cm,angle=0 }}
  \caption{Averaged EIS count rates (DN/s per 1\arcsec) in the QS areas,
corrected for the LW long-term decrease in sensitivity.}
 \label{fig:quiet_int_corr}
\end{figure}

\begin{figure}[!htbp]
\centerline{\epsfig{file=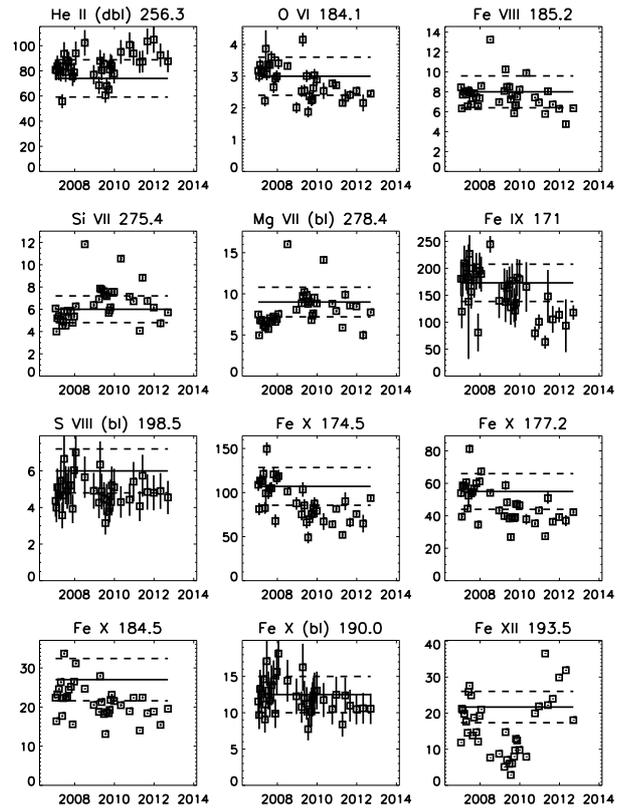, width=8.cm,angle=0 }}
  \caption{Averaged EIS QS radiances as in  Fig.~\ref{fig:quiet_rad_std},
but obtained with the present calibration.}
 \label{fig:quiet_rad_gdz}
\end{figure}

 \begin{figure}[!htbp]
\centerline{\epsfig{file=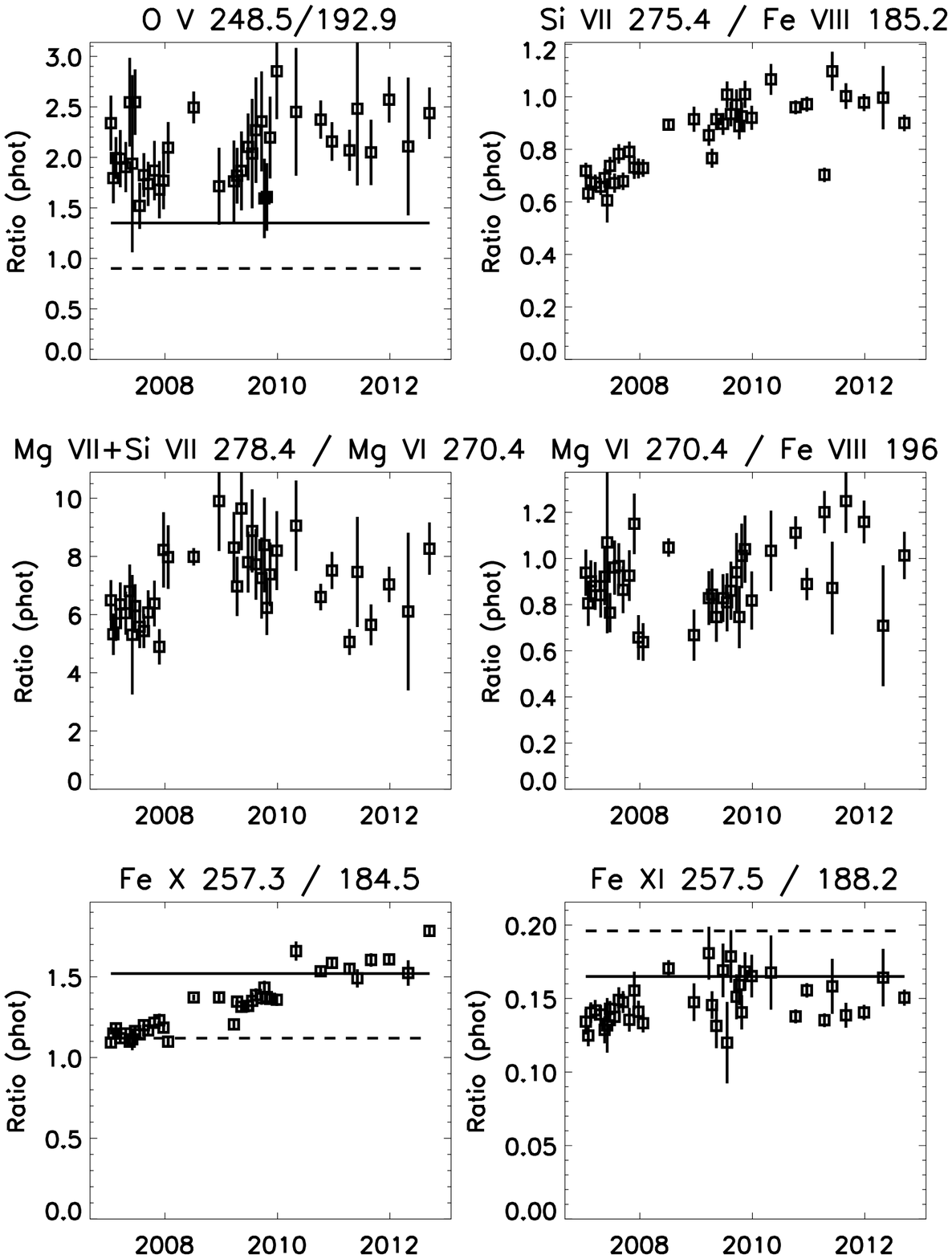, width=8.cm,angle=0 }}
\centerline{\epsfig{file=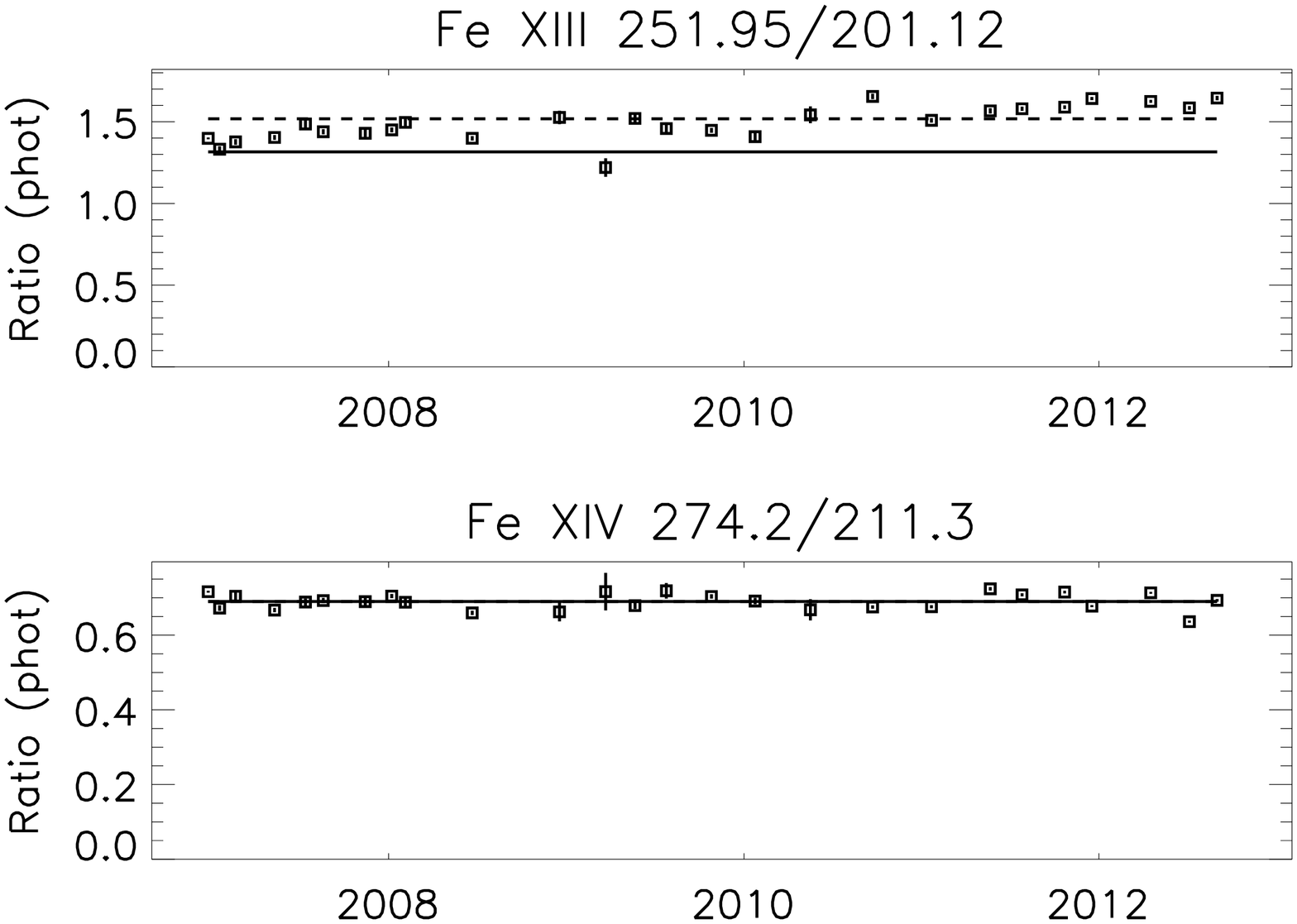, width=6.cm,angle=90 }}
  \caption{Ratios of a few QS and AR line radiances (photon units)
in the SW and LW channels, obtained with  the present first-order
calibration. Bars indicate 
the predicted  values. }
 \label{fig:ratios_corr}
\end{figure}

We have applied the present  first-order calibration  to several
cases spread over the years. We have found good overall agreement
between observations and theory. Some examples are given below.
We start by showing in 
 Fig.~\ref{fig:quiet_int_corr} the QS count rates in a few LW lines,
corrected for the LW long-term decrease in sensitivity.
 The figure   shows that we obtain  relatively constant values in the cool 
 lines, most notably the \ion{He}{ii} 256.3~\AA\ line (which has been deblended),
but also the \ion{O}{iv}, \ion{O}{v}, and \ion{Mg}{vi} lines. 
This confirms the reliability of the long-term correction.
Some residual variations and a large scatter are present, but they 
are due to solar variability. We recall that the synoptic observations
were carried out on small fields of view and are therefore not
ideal to carry out a proper study of the solar radiances.

Fig.~\ref{fig:quiet_rad_gdz} shows the 
 EIS QS calibrated radiances of the same lines shown in 
 Fig.~\ref{fig:quiet_rad_std} (those for which we found QS historical values),
 obtained with the present first-order calibration, i.e. both 
new effective areas and long-term correction. 
There is clearly a very good agreement (to with a relative $\pm$20\%) with the 
previous measurements, during the first two years of the mission,
when the Sun was quiet.

 Fig.~\ref{fig:ratios_corr} shows the ratios of a few QS and AR 
line radiances (photon units)
in the SW and LW channels, obtained with  the present first-order
calibration. The \ion{O}{v} ratio shows an overall constancy.
The correction based on the \ion{Fe}{xiv} ratio  reproduces
extremely well the \ion{Fe}{xiii} ratio and 
reasonably well the \ion{Fe}{xi} 257.5 / 188.2~\AA\ one. 
The  increase in the \ion{Fe}{x} 257.3/184.5~\AA\ is puzzling.
If density variations are neglected, the trend would imply 
a decrease in the temperature, which is contrary to what is 
observed. The increase could partly be
ascribed to a decrease in the effective area at 184.5~\AA\ not taken into 
account. 
The \ion{Si}{vii} 275.4~\AA / \ion{Fe}{viii} 185.2~\AA\ corrected ratio
shows a residual increase over time. We discuss this in the following section.

\subsection{The \ion{Si}{vii} vs. \ion{Fe}{viii}  issue}

We recall that the  \ion{Si}{vii} 275.3~\AA\ vs. \ion{Fe}{viii} 185.2~\AA\ 
 ratio was used by \cite{kamio_mariska:12} and \cite{mariska:12_eis_cal} to argue
that both LW and LW channels had a similar degradation.
There are two problems in using this ratio, however. First, the 
\ion{Si}{vii} and \ion{Fe}{viii} are probably not formed in the 
same spatial regions. This is because the temperature of formation of \ion{Si}{vii} 
is different  than that of \ion{Fe}{viii}
(note though that the \ion{Fe}{viii} formation temperature  has been changing 
significantly over the years, as calculations and measurements of the 
ionisation and recombination rates have been  improved).

The second problem is that some variations in the iron vs. silicon 
abundances, although not expected on a large scale, could be happening.
Both effects might be the reason for the large scatter in the 
ratios of the \ion{Si}{vii} and \ion{Fe}{viii} lines, shown in 
 Fig.~\ref{fig:ratios_sw_lw}.
Note that the scatter is not limited to the two lines, because
all the \ion{Fe}{viii} lines and all the  \ion{Si}{vii} lines have 
similar trends, i.e. their ratios are relatively constant, as shown 
in the previous figures.

It is also clear that other ratios involving 
the \ion{Mg}{vi} and \ion{Fe}{viii} lines behave in a different way,
as shown in Fig.~12 with the 
 \ion{Mg}{vi} 270.4/ \ion{Fe}{viii} 196~\AA.
  There is still scatter but the ratios are fairly
constant once the present long-term correction is applied.
We have checked that all the \ion{Mg}{vi} lines in the LW channel behave in the same
way. The puzzling issue is that the \ion{Mg}{vi} lines are close
in wavelength to the \ion{Si}{vii} lines. 

The fact that the \ion{Mg}{vi} vs. \ion{Fe}{viii} ratios 
show less scatter and a different trend over time, compared to the 
\ion{Si}{vii} vs. \ion{Fe}{viii} ratios, might be due to  
the fact that the TR components of  \ion{Mg}{vi} and \ion{Fe}{viii}
originate from the same temperatures/spatial regions.
This is further discussed in the Appendix.

\subsection{Observations in 2012}

Several recent observations in 2012 were then analysed, to see if this
first order calibration is sufficient to remove the main problems.
We present here only the results of the analysis of
 an off-limb AR observation of 2012 Apr 16 and a 
flare observation of 2012 Mar 9. 
A preliminary analysis of the 2012 Apr 16 observation indicates that large 
departures in the shapes of the effective areas have not 
occurred, with the exception of the shorter wavelengths of the 
SW channel, where the sensitivity has further decreased.
This is all consistent  with the behaviour of the line ratios discussed previously.

The 2012 Mar 9 observation is particularly important since it is the first and
only EIS full-spectral observation of a medium-size M-class flare.
A detailed description of this observation is presented in a separate paper.
Here, we focus on the \ion{Fe}{xvii} and \ion{Fe}{xxiv} lines.

Extra care was needed to select the best spectra for either 
\ion{Fe}{xvii} or \ion{Fe}{xxiv}. 
For example, various lines were saturated in many exposures. 
The \ion{Fe}{xxiv} lines during the impulsive phase  have 
large blue-shifted components which complicate the analysis.
A spectrum for \ion{Fe}{xvii} and one for \ion{Fe}{xxiv} was obtained.
In the first spectrum, the \ion{Fe}{xvii} lines are so bright that 
significant blending should not be present. The \ion{Fe}{xvii} lines 
are very useful to check the calibration because they fall across
most LW wavelengths.
Discrepancies of over a factor 
of two were found in the \ion{Fe}{xvii} 204.7 / 254.9~\AA\
 and \ion{Fe}{xxiv} 192 / 255.1~\AA\ ratios using the ground calibration.
The first-order calibration presented above removes these main discrepancies.

\begin{figure}[!htbp]
\centerline{\epsfig{file=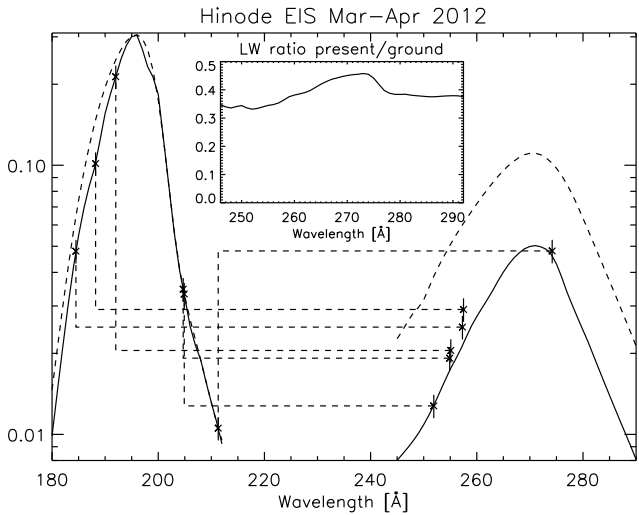, width=7.cm,angle=90}}
\caption{Effective areas for the EIS channels obtained with the 
first-order calibration for the period Mar--Apr 2012.
The full lines indicate the proposed values, while the dashed ones indicate those
from the ground calibration.
The ratios of the effective areas as obtained from 
the off-limb AR observation of 2012 Apr 16 and the 
flare observation of 2012 Mar 9 are overplotted, showing 
good agreement. 
}
 \label{fig:eff_apr_2012} 
\end{figure}

The March and April 2012 line ratios have been 
combined  in  Fig.~\ref{fig:eff_apr_2012} by applying the 
first-order correction. The long-term correction indicates a 
decrease of about a factor of two in the relative LW/SW
sensitivity for this period. The very reliable \ion{Fe}{xvii} and \ion{Fe}{xxiv} 
ratios show very good agreement  with the present correction
(see points at 254.9, 255~\AA).
In the \ion{Fe}{xxiv} spectrum, 
the \ion{Fe}{xxiv} 192 / 255.1~\AA\ ratio is about 4 (photon units)
if the ground calibration is used,
instead of the expected 1.85.
The present first-order calibration brings the ratio down to 
a value of 1.6, i.e. within a reasonable 15\% the expected value. 
The  agreement in the ratios involving lines short-ward of 190~\AA\ can 
be improved if a wavelength-dependent correction for the SW channel is introduced.

\begin{figure}[!htbp]
\centerline{\epsfig{file=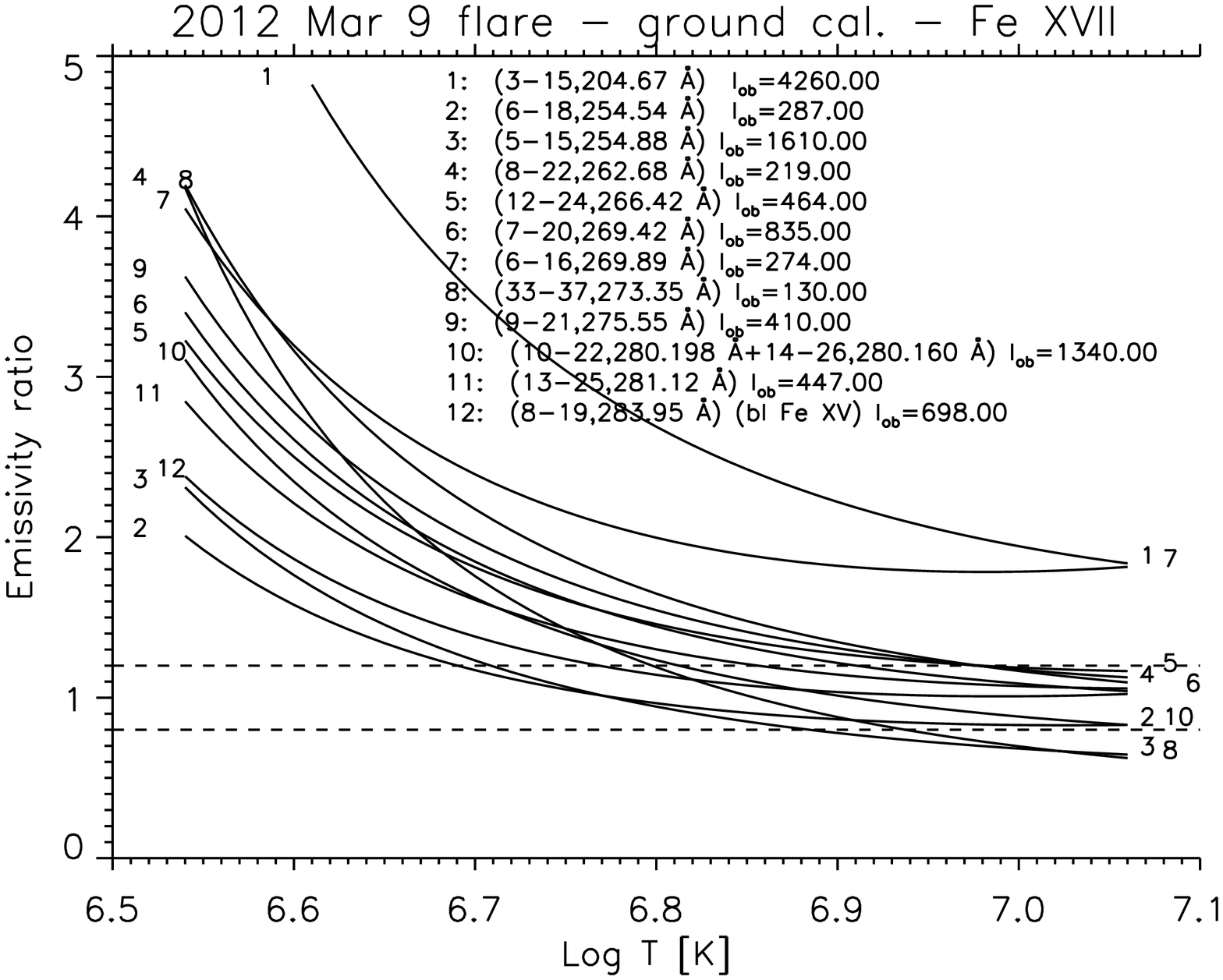, width=7.cm,angle=90 }}
\centerline{\epsfig{file=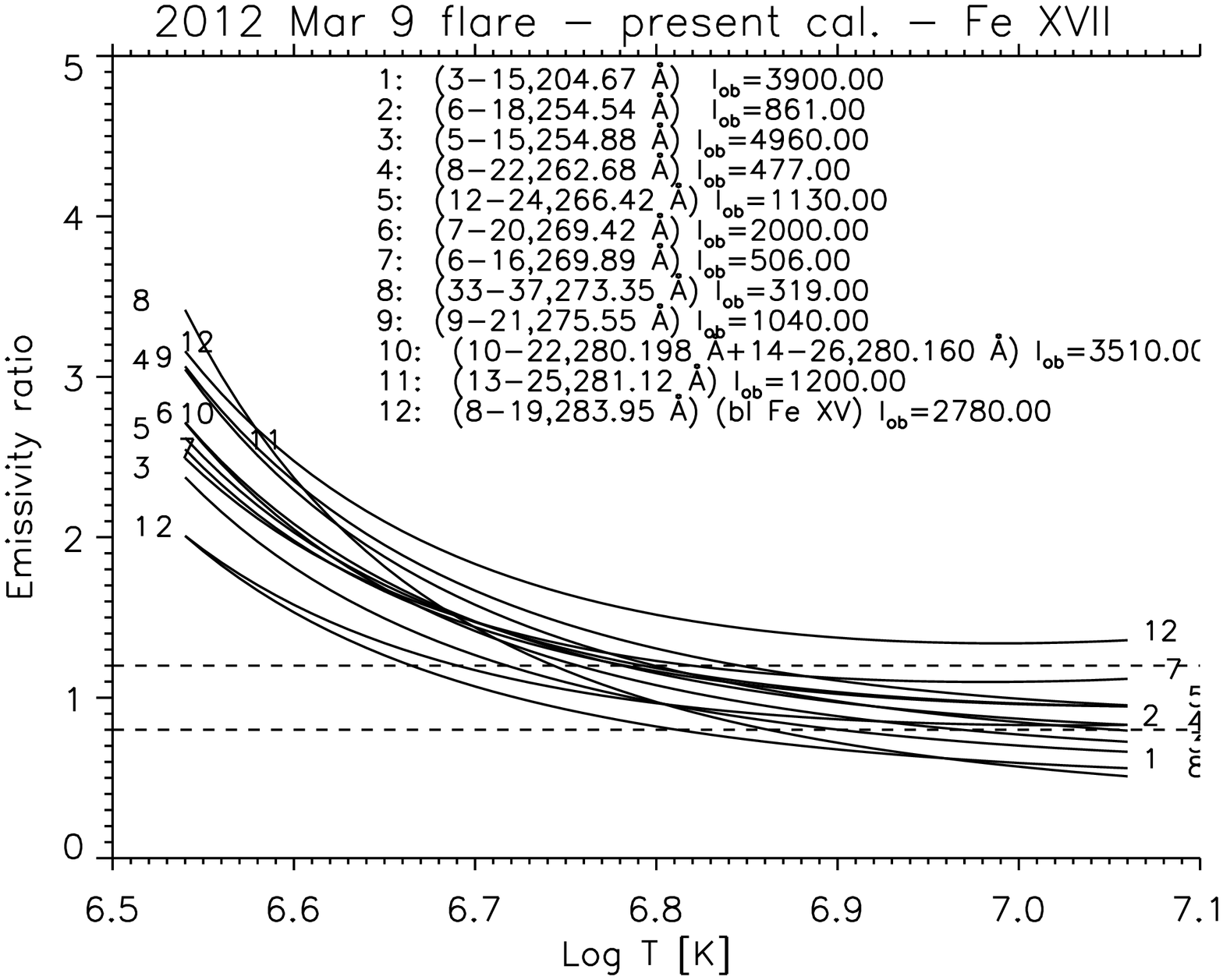, width=7.cm,angle=90 }}
  \caption{Emissivity ratio  curves  relative to  the 
main \ion{Fe}{xvii} EUV transitions observed  by Hinode  EIS on 2012 Mar 9.
$I_{\rm ob}$ is the calibrated observed intensity in photons 
cm$^{-2}$ s$^{-1}$ arcsec$^{-2}$.
The top plot shows the results with the current ground calibration,
while the bottom one with the present first-order correction. 
}
  \label{fig:fe_17_flare}
\end{figure}

Fig.~\ref{fig:fe_17_flare} shows the 
emissivity ratio  curves  relative to  the 
main \ion{Fe}{xvii} lines.
These  emissivity ratio  curves are obtained by dividing 
 the observed intensity $I_{\rm ob}$ of a line  by its emissivity:
\begin{equation}
F_{ji}= { I_{\rm ob} \; N_{\rm e} \; C \over N_j(N_{\rm e}, T_{\rm e})  \;A_{ji}}
\end{equation}
 calculated at a fixed electron  density $N_{\rm e}$
and  plotted  as a function of the  temperature $T_{\rm e}$
(see \citealt{delzanna_etal:04_fe_10} for details).
 The scaling constant $C$ is chosen so that the 254.9~\AA\ curve
is  close to unity.

 The emissivity ratio  curves show the large discrepancy 
(more than a factor of two) in the 
\ion{Fe}{xvii} 204.7 / 254.9~\AA\ branching ratio, when the 
current ground calibration is adopted (top plot).
The present first-order calibration brings the ratio within 
an excellent 10\%.
With the new calibration, there is very good agreement (to within a relative 20\%)
between theory and observation for all the \ion{Fe}{xvii} lines,
a remarkable result which further  confirms the present calibration. 
The only line with a significant departure is very close to the
\ion{Fe}{xv} 284.1~\AA\ line and is difficult to measure it 
accurately. On the other hand, large discrepancies 
are present if the ground calibration is adopted.
All the \ion{Fe}{xvii} lines are very strong, but some small 
blending is expected to be present. Further refinements 
will need an in-depth analysis that is left to a separate paper.
 The plot also shows that the ratio of the \ion{Fe}{xvii}
273.35~\AA\ line with any of the other 
 lines can potentially be used to measure electron temperatures for flares.

\section{Conclusions}

This preliminary assessment of the in-flight degradation of the 
EIS instrument based on line radiances and ratios
  shows a consistent pattern that is fundamentally
different from what has been assumed so far.

The observations show a clear  degradation of the 
long-wavelength  channel, compared to the short-wavelength one.
The responsivity of the LW channel, at its shorter wavelengths,
 in December 2006 was already lower (by 20--30\%, see inset in the middle plot
in Fig.~8) than what was measured on the ground.
It continued to decrease significantly until 2010, when 
radiances in LW lines became underestimated by a factor of 2.
On the other hand, QS radiances in SW lines do not 
show any indication of a major degradation for the SW channel,
with the exception  of its shorter wavelengths.

Overall, the shapes of the effective areas  in their central regions
are close  to  those measured on the ground.
Significant departures (30--50\%) are  present towards the  edges of both 
SW and LW channels, however.
Such small departures are not surprising, especially considering the large discrepancies 
 (up to 50\%) within  the ground measurements, discussed previously. 

The present  LW long-term correction  brings  the main LW/SW 
line ratios to become constant to within a reasonable 20\%, 
and  produces  relatively constant QS radiances in the LW cool lines,
most notably the strongest line, \ion{He}{ii} 256.3~\AA.

The present new calibration also removes a number of peculiar features in terms of 
emission measures and elemental abundances that 
we have encountered.
A significant number of studies such as those concerning 
emission measures combine the use of EIS cooler lines 
(e.g. \ion{Fe}{viii} -- \ion{Fe}{xiii}) from the 
SW channel with the hotter ones (\ion{Fe}{xiv}--\ion{Fe}{xvi})
from the LW channel. Such studies should be revised by taking into 
account the decrease in the  responsivity of the LW channel.
 
The EIS instrument has  performed reasonably well,
with only a factor of two clear degradation in one of the channels
within the first two years of the mission.
Note that significant degradations (order of magnitude) are  very common, even in 
  instruments recently built, especially if they have a front filter.
For example, already at first light, the  SDO EVE MEGS-B showed a factor of 10
drop in sensitivity, and significant decreases are still occurring
at selected wavelengths.
Such degradations are measured by using different filters, some only 
occasionally for calibration purposes. 
Different front filters show very different attenuations. 
The SDO EVE ESP instrument has an Al filter, and shows similar 
attenuation as  the MEGS-A2 instrument, which has an Al-Ge-C filter. 
The MEGS-A2 shows a clear wavelength dependence, with lines at 
longer wavelengths more attenuated. After two years, lines around 
190~\AA\ were attenuated by 30\%, while those at 284~\AA\ by  50\%.
This is interpreted as carbon deposition in the front filter.
The LYRA instruments on-board PROBA-2 also suffered large degradations,
known to be caused  by  contamination (probably molecular) 
on the front optical filters.

Other instruments such as the SOHO CDS performed much better, with a degradation 
of only a factor of 2 over 13 years
\citep{delzanna_etal:10_cdscal}, perhaps because of the 
rigorous cleanliness program and/or because of the lack of an entrance filter.
Such calibration issues were discussed at the 
on-orbit degradation workshop that took place in the Solar
Terrestrial Centre of Excellence (STCE, Royal Observatory of Belgium) 
in Brussels on May 3, 2012. A summary  is given in \citep{aly_etal:12}.
As a follow-up, further meetings will be hosted by STCE, 
to check the inter-calibrations of the various instruments. 

Out-gassing and contamination of the entrance filter could partly
be responsible for 
 the observed degradation in the EIS LW. Deposition of carbon compounds 
 would cause an enhanced degradation in the LW channel,
compared to the SW one, as observed in the SDO EVE MEGS-A2.
However, they would also cause some degradation in the SW channel.
We have seen some evidence of degradation at the SW shorter wavelengths
but not at the longer ones, which is not expected.
The present calibration  is a contribution to the 
 on-going effort within the EIS team to understand 
the  degradation of the instrument and to provide the best possible correction.

A preliminary comparison between EIS and EVE data since May 2010 
has shown very similar results to those presented here 
(I.Ugarte Urra and H.Warren, priv. comm.). 
All the results shown in \cite{mariska:12_eis_cal} are also in agreement 
with those shown here, although the interpretation is very different.
\cite{mariska:12_eis_cal} assumes that there is something odd 
about the \ion{He}{ii} 256.3~\AA\ line, 
and that the LW/SW calibration can be checked using the 
\ion{Si}{vii} 275.4~\AA\ vs. \ion{Fe}{viii} 185~\AA\ ratio.
We have seen the various problems associated with this ratio.
The suggestion from \cite{mariska:12_eis_cal}
 does not explain the factor of two problems with the 
\ion{Fe}{xi}, \ion{Fe}{xiii}, \ion{Fe}{xiv}, \ion{Fe}{xvii}, \ion{Fe}{xxiv}
lines observed in the SW and LW channels in the data after 2010.
The present calibration resolves all the discrepancies, but leaves
somewhat unexplained the behavior of some lines. 
Further work can be done on the lines of the present study, however what 
is needed is a calibration rocket.
In the near future, a new EUNIS rocket should be launched,
which will provide important information on the current EIS calibration.

\begin{acknowledgements}
Useful discussions and comments from various members of the EIS team
are acknowledged, in particular from   J. Mariska, 
T. Watanabe, H. Hara., H. Warren. 
A special thank goes to the referee, P. Young, for detailed discussions and 
suggestions on how to improve the analysis. \\

Support from  STFC  is  acknowledged. 
The work of the UK APAP Network was funded by the UK
STFC under grant No. PP/E001254/1 with the University of Strathclyde.
CHIANTI is a collaborative
project involving researchers  at  the
Universities of  Cambridge (UK), George Mason, Michigan (USA).
The excellent Hinode Science Data Centre Europe was used to 
search the EIS database. \\
Hinode is a Japanese mission developed and launched by ISAS/JAXA, with
NAOJ as domestic partner and NASA and STFC (UK) as international
partners. It is operated by these agencies in co-operation with ESA
and NSC (Norway). 

\end{acknowledgements}

\bibliographystyle{aa}

\bibliography{../bib}



\appendix

\section{List of the observations }

\def\baselinestretch{1.}

\begin{table*}[!htbp]
\caption{QS observations used for the EIS calibration.}
\begin{center} 
\footnotesize
\begin{tabular}{@{}llllll@{}}

 \hline\hline \noalign{\smallskip}
File & Raster         & Slit (\arcsec)  & Exp. (s) & FOV (\arcsec) &  \\ 
\hline \noalign{\smallskip}

20061223\_161013 & HPW001\_FULLCCD\_RAST & 1 & 90 & 128$\times$128 & \\


20070114\_220819 & SYNOP001\_slit & 1 &  90 & 1$\times$256  & \\
20070130\_111912 & HPW001\_FULLCCD\_RAST & 1 & 90 & 128$\times$128 & \\
20070216\_112350 & SYNOP001\_slit & 1 &  90 & 1$\times$256  & \\
20070316\_180127 &  SYNOP001\_slit & 1 &  90 & 1$\times$256  & \\
20070421\_000705 & SYNOP001\_slit & 1 &  90 & 1$\times$256  & \\

20070517\_000450 & SYNOP001\_slit & 1 &  90 & 1$\times$256  & \\

20070602\_131520 & HPW008\_FULLCCD\_RAST &  1 & 25 & 128$\times$128 & \\
20070620\_180835 & SYNOP001\_slit & 1 &  90 & 1$\times$256  & \\
20070720\_110822 & SYNOP001\_slit & 1 &  90 & 1$\times$256  & \\

20070817\_062935 & SYNOP001\_slit & 1 &  90 & 1$\times$256  & \\ 
20070913\_175836 & SYNOP001\_slit & 1 &  90 & 1$\times$256  & \\ 
20071024\_061835 & SYNOP001\_slit & 1 &  90 & 1$\times$256  & \\ 
20071125\_104656 & SYNOP001\_slit & 1 &  90 & 1$\times$256  & \\ 
20071222\_111205 & SYNOP001\_slit & 1 &  90 & 1$\times$256  & \\ 

20080121\_160213 & SYNOP002\_FULLCCD & 1 & 90 & 128$\times$184 & \\

20080705\_112034 & FELDMAN\_QSCH\_ATLASv1 & 2 & 120 & 24$\times$304 & \\


\\

20081217\_110519 & HPW001\_FULLCCD\_RAST & 1 & 90 & 128$\times$128 & \\


20090323\_174230 &  SYNOP001\_slit & 1 &  90 & 1$\times$256  & \\ 
20090413\_175041 & HPW001\_FULLCCD\_RAST & 1 & 90 & 128$\times$128 & \\

20090511\_180929 & SYNOP001\_slit & 1 &  90 & 1$\times$256  & \\ 
20090623\_182812 & SYNOP001\_slit & 1 &  90 & 1$\times$256  & \\ 
20090720\_060035 & SYNOP001\_slit & 1 &  90 & 1$\times$256  & \\ 
20090813\_180429 & SYNOP001\_slit & 1 &  90 & 1$\times$256  & \\ 
20090919\_180557 & SYNOP001\_slit & 1 &  90 & 1$\times$256  & \\ 

20091007\_120219 & HPW001\_FULLCCD\_RAST & 1 & 90 & 128$\times$128 & \\%
20091023\_060550 & SYNOP001\_slit & 1 &  90 & 1$\times$256  & \\ 
20091113\_180529 & SYNOP001\_slit & 1 &  90 & 1$\times$256  & \\ 
20091227\_063535 & SYNOP001\_slit & 1 &  90 & 1$\times$256  & \\ 

20100501\_054013 &  Atlas\_30 & 2 & 30 & 120$\times$160 & \\
20101008\_101526 & Atlas\_120 & 2 & 120 & 120$\times$160 & \\
20101220\_050526 & Atlas\_120 & 2 & 120 & 120$\times$160 & \\ 
20110413\_132033 & Atlas\_120 & 2 & 120 & 120$\times$160 & \\ 

20110603\_113020 & Atlas\_060x512\_60s & 1 & 60 & 60$\times$512  & \\
20110831\_054534 &  Atlas\_60 & 2 & 60 & 120$\times$160 & \\
20111226\_181940 & Atlas\_120 & 2 & 120 & 120$\times$160 & \\ %
20120428\_151319 & Atlas\_060x512\_60s & 1 & 60 & 60$\times$512  & \\
20120913\_182534 & Atlas\_120 & 2 & 120 & 120$\times$160 & \\ %

\noalign{\smallskip}\hline  
\end{tabular}
\normalsize
\tablefoot{The columns indicate the file name (date and UT time),
the EIS raster acronym, the slit used, the exposure time and 
the field of view (FOV) of the observation. }
\end{center}
\label{tab:list_files_qs}
\end{table*}

\begin{table*}[!htbp]
\caption{AR observations used for the EIS calibration.}
\begin{center} 
\footnotesize
\begin{tabular}{@{}llllll@{}}

 \hline\hline \noalign{\smallskip}
File & Raster         & Slit (\arcsec)  & Exp. (s) & FOV (\arcsec) &  \\ 
\hline \noalign{\smallskip}

20061225\_221058 &  SYNOP001\_slit & 1 &  90 & 1$\times$256  & \\ %


20061225\_225013 & HPW001\_FULLCCD\_RAST & 1 & 90 & 128$\times$128 & \\%

20070118\_120435 &  SYNOP001\_slit & 1 &  90 & 1$\times$256  & \\ %
20070210\_000119 &  SYNOP001\_slit & 1 &  90 & 1$\times$256  & \\ %
20070220\_054028 &  SYNOP001\_slit & 1 &  90 & 1$\times$256  & \\ %
20070322\_125559 &  SYNOP001\_slit & 1 &  90 & 1$\times$256  & \\ %
20070423\_184443 &  SYNOP001\_slit & 1 &  90 & 1$\times$256  & \\ %
20070511\_105544 &  SYNOP001\_slit & 1 &  90 & 1$\times$256  & \\ %
20070519\_180450 &  SYNOP001\_slit & 1 &  90 & 1$\times$256  & \\ %
20070607\_181319 &  SYNOP001\_slit & 1 &  90 & 1$\times$256  & \\ %
20070630\_175020 &  SYNOP001\_slit & 1 &  90 & 1$\times$256  & \\ %
20070630\_175235 &  SYNOP001\_slit & 1 &  90 & 1$\times$256  & \\ %
20070630\_175451 &  SYNOP001\_slit & 1 &  90 & 1$\times$256  & \\ %
20070630\_175706 &  SYNOP001\_slit & 1 &  90 & 1$\times$256  & \\ %
20070714\_000949 &  SYNOP001\_slit & 1 &  90 & 1$\times$256  & \\ %

20070819\_133227 & HPW001\_FULLCCD\_RAST & 1 & 90 & 128$\times$128 & \\%

20070929\_102529 &  SYNOP001\_slit & 1 &  90 & 1$\times$256  & \\ %
20071020\_021049 &  SYNOP001\_slit & 1 &  90 & 1$\times$256  & \\ %
20071114\_000707 & SYNOP001\_slit & 1 &  90 & 1$\times$256  & \\ %
20071206\_175549 &  SYNOP001\_slit & 1 &  90 & 1$\times$256  & \\ %
20080107\_101448 &  SYNOP001\_slit & 1 &  90 & 1$\times$256  & \\ %
20080118\_103435 &  SYNOP001\_slit & 1 &  90 & 1$\times$256  & \\ %
20080204\_104700 &  SYNOP001\_slit & 1 &  90 & 1$\times$256  & \\ %

20080620\_230339 & HPW001\_FULLCCD\_RAST & 1 & 90 & 128$\times$128 & \\%
\\

20081217\_110519 & HPW001\_FULLCCD\_RAST & 1 & 90 & 128$\times$128 & \\%


20090322\_060630 &  SYNOP001\_slit & 1 &  90 & 1$\times$256  & \\ %

20090521\_180529 & SYNOP001\_slit & 1 &  90 & 1$\times$256  & \\ %

20090725\_055049 & SYNOP001\_slit & 1 &  90 & 1$\times$256  & \\ %

20091025\_214625 & HPW008\_FULLCCD\_RAST &  1 & 25 & 128$\times$128 & \\

20100123\_171532 & SYNOP001\_slit & 1 &  30 & 1$\times$256  & \\ %

20100517\_135741 & Atlas\_60 & 2 & 60 & 120$\times$160 & \\
20100922\_112633 & Atlas\_60 & 2 & 60 & 120$\times$160 & \\

20110121\_123757 & Atlas\_60 & 2 & 60 & 120$\times$160 & \\

20110522\_103354 & Atlas\_60 & 2 & 60 & 120$\times$160 & \\

20110726\_175935  & Atlas\_60 & 2 & 60 & 120$\times$160 & \\
20111022\_100543 & Atlas\_60 & 2 & 60 & 120$\times$160 & \\

20111217\_125856 & Atlas\_60 & 2 & 60 & 120$\times$160 & \\
20120416\_124033 & Atlas\_60 & 2 & 60 & 120$\times$160 & \\
20120704\_223657 & Atlas\_60 & 2 & 60 & 120$\times$160 & \\

20120830\_234044  & Atlas\_60 & 2 & 60 & 120$\times$160 & \\

\noalign{\smallskip}\hline  
\end{tabular}
\normalsize
\end{center}
\label{tab:list_files_ar}
\end{table*}

\section{Effective areas and long-term correction}

In what follows, we provide some IDL commands to obtain the 
effective areas and the long-term correction within SolarSoft.

\begin{verbatim}

gdz_sw=[165, 171, 174.5, 177.2, 178.1, 180.4, 182.2, $
 184.5, 185.2, 186.9,  188.3, 190,  $
         192.4, 192.8, 193.5, 194.7, 195.1, $
         196.6, 197.4, 200,  201.1,  202.,  $
         202.7, 204.9, 208., 209.9, 211.3]
 
gdz_eff_sw=$
[0.000174973/1.5,0.000255772/1.5,0.00158207/1.5, $
0.00476608/1.55, 0.00705735 /1.5, 0.0168637/1.45,$
0.0316499/1.4, 0.0647319/1.35, 0.0779082/1.35,$
0.115240/1.4,  0.150199/1.45, 0.194897/1.25, $
0.255993/1.13, 0.264945/1.1, 0.279607/1.05, $
0.298884/1.02,  0.302737*1., 0.301859/1.05,$
0.287675/1.15, 0.174608*1.05,  0.119586/1.0, $
 0.0838537 /1., 0.0635698/1.,  0.0332376/1.0 ,$ 
0.0189209/1., 0.0133581/1., 0.0105513/1.]

gdz_lw=[245., 252., 255,  257., 259, $
        263. , 265., 268.,  270., $
        272. , 274. , 277.,    281., 286., 292]

gdz_eff_lw=$
[0.022673*0.8, 0.03908*0.75, 0.05065*0.78, $
0.0588*0.8, 0.06738*0.85, 0.0861*0.9,$
0.09551*0.95, 0.106984*1.0, 0.110764*1.02, $
0.10944*1.03,  0.1026*1.03, 0.084775*0.9, $
0.05718*0.87, 0.0333*0.85, 0.01679*0.85]/1.1

; the effective area at a SW wavelength (Angstroms) 
;w1 is obtained as 
; eff=interpol(gdz_eff_sw , gdz_sw, w1),/spline)
; the effective area at a LW wavelength (Angstroms)
; w2 is obtained as 
; eff= interpol(gdz_eff_lw , gdz_lw, w2),/spline)

;--------------------------------------------------

Xtime_ref=ANYTIM2TAI('22-Sept-2006 21:36:00.000')
coeff1=[1.0326230,  -5.2495791e-09,   1.2055185e-17]

; date is any observation time.
Xtime=ANYTIM2TAI(date)

; degr gives the degradation
degr=poly(Xtime - Xtime_ref, coeff1)  
 

\end{verbatim}

\section{The \ion{Si}{vii} vs. \ion{Fe}{viii}  problem}

\begin{figure}[!htbp]
\centerline{\epsfig{file=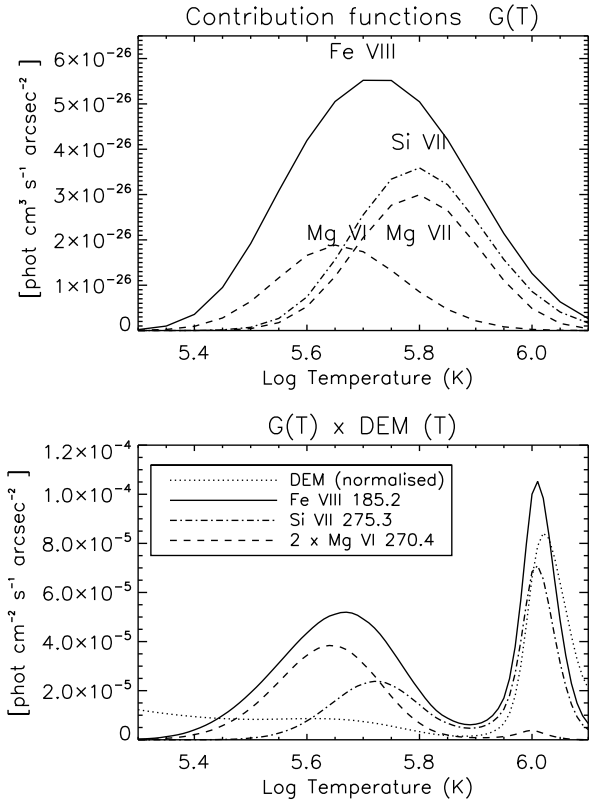, width=8.cm,angle=0 }}
  \caption{Top: contribution functions $G(T)$ of a selection of TR lines
observed by Hinode/EIS, using CHIANTI v.7.1 data. 
Bottom: the same $G(T)$ values multiplied for a QS DEM. 
Note that the  $G(T) \times DEM(T)$ of  \ion{Mg}{vi} was increased by a 
factor of two.}
 \label{fig:this_goft}
\end{figure}

We have calculated the contribution functions $G(T)$ of a selection of TR lines
observed by Hinode/EIS, using CHIANTI v.7.1 data \citep{landi_etal:11_chianti_v7}, in particular the 
new ion abundances (in ionization equilibrium). 
They are shown in Fig.~\ref{fig:this_goft} (top).
Any small differences in the contribution functions of two lines
in  the transition region, where there is a steep variation in 
temperature, can have a large effect (see, for example, the large effects on 
active region loops \citealt{delzanna:03}).

To estimate the effects that a steep variation in temperature can have, 
we have taken a QS DEM obtained from the on-disk EIS observations 
of 2006 Dec 23, and folded it with the $G(T)$ of a few main EIS lines,
as an example. They are shown in Fig.~\ref{fig:this_goft} (bottom).
It is clear that the emission in the  \ion{Fe}{viii} and 
 \ion{Si}{vii} has two main components, one coronal (1 MK)
and one in the lower transition region. 
It is also clear that the TR component of \ion{Fe}{viii} should have a 
similar response to that of \ion{Mg}{vi}, and not \ion{Si}{vii}.
It is fair to say though that for both \ion{Fe}{viii} and 
 \ion{Si}{vii} a significant contribution is predicted to originate
from the coronal component. This is of course assuming that a 
continuous distribution of plasma between TR temperatures and 
the corona exists.
The TR \ion{Si}{vii} component 
accounts for almost 30\% of the intensity of the line
 in the log T[K]=5.2--5.9 range, while the \ion{Fe}{viii} TR component 
accounts for over 40\%. \ion{Mg}{vi} is clearly a purely TR line.

\end{document}